\documentclass[fleqn,usenatbib]{mnras}

\usepackage{newtxtext,newtxmath}

\usepackage[T1]{fontenc}

\DeclareRobustCommand{\VAN}[3]{#2}
\let\VANthebibliography\thebibliography
\def\thebibliography{\DeclareRobustCommand{\VAN}[3]{##3}\VANthebibliography}

\usepackage{graphicx}	
\usepackage{amsmath}	

\usepackage[dvipsnames]{xcolor}
\usepackage{tikz,hyperref}

\definecolor{lime}{HTML}{A6CE39}
\DeclareRobustCommand{\orcidicon}{
	\begin{tikzpicture}
	\draw[lime, fill=lime] (0,0) 
	circle [radius=0.14] 
	node[white] {{\fontfamily{qag}\selectfont \tiny ID}};
	\draw[white, fill=white] (-0.0625,0.000) 
	circle [radius=0.007];
	\end{tikzpicture}
	\hspace{-2mm}
}

\foreach \x in {A, ..., Z}{\expandafter\xdef\csname orcid\x\endcsname{\noexpand\href{https://orcid.org/\csname orcidauthor\x\endcsname}
			{\noexpand\orcidicon}}
}

\title[Irregular grains]{Photoionization modelling of circumstellar nebulae using irregular grains}

\author[Jim\'{e}nez-Hern\'{a}ndez et al.]{P.\,Jim\'{e}nez-Hern\'{a}ndez\thanks{E-mail: pajimenez@astro.unam.mx}$^{1\orcidA}$, S.\,J.\,Arthur$^{2\orcidB}$, Daniel\,Guirado$^{3}$, Olga\,Muñoz$^{3\orcidD}$, Julia\,Martikainen$^{3\orcidE}$, \newauthor{L.\,Sabin$^{1\orcidF}$} and William\,J. Henney$^{2\orcidG}$.\\
$^{1}$Universidad Nacional Aut\'onoma de M\'exico. Instituto de Astronom\'{\i}a. A.P. 106, Ensenada 22800, B.C., Mexico.\\
$^{2}$Instituto de Radioastronom\'{\i}a y Astrof\'{\i}sica, Universidad Nacional Aut\'onoma de México, Antigua Carretera a P\'atzcuaro \#8701, Ex-Hda.\ San Jos\'e de la Huerta,\\ Morelia, Michoac\'an, C.P. 58089, Mexico.\\
$^{3}$Instituto de Astrof\'{\i}sica de Andaluc\'{\i}a (IAA-CSIC), Glorieta de la Astronom\'{\i}a s/n, E-18008 Granada, Spain.
}

\date{Accepted XXX. Received YYY; in original form ZZZ}

\pubyear{\the\year{}}

\begin{document}
\label{firstpage}
\pagerange{\pageref{firstpage}--\pageref{lastpage}}
\maketitle

\begin{abstract}
We study the effects of using the optical properties of irregular hexahedral grains in photoionization models of circumstellar nebulae around evolved stars. Dust opacities for the irregular grains were obtained from the scattering properties available in the TAMUdust2020 database and these were implemented in the spectral synthesis code \textsc{cloudy}. A sample of photoionization models that use opacities from both spherical and irregular hexahedral grains across a standard MRN size distribution (0.005–0.25 $\mu$m) was produced. We consider the optical properties of graphite, amorphous carbon and silicate dust grains and find that differences between the model nebula continua calculated using spherical and irregular dust grains increase with the grain size, especially for graphite. In particular, we find that the luminosities at the infrared peak for the hexahedral grain models can be up to 60\% higher than those from the equivalent spherical grain models for the largest grains. This result suggests that traditional spherical grain assumptions may lead to an overestimate of the dust mass in photoionized nebulae.
\end{abstract}




\section{Introduction}
\label{sec:intro}

The cool outflows of AGB stars and red supergiants are important sites for dust formation \citep{2003dge..conf.....W}. Dust grains are expelled in the slow, strong stellar winds and end up in the circumstellar media and nebulae that surround the hot, later stages of stellar evolution, where they reprocess the stellar radiation and are responsible for the nebular emission at mid-infrared (MIR) to far-infrared (FIR) wavelengths. Photoionization modelling of planetary nebulae (PN) and Wolf-Rayet nebulae (WRN) enables us to estimate the mass and deduce the chemical composition of the grain material, which are important for understanding stellar evolution and the chemical enrichment of the interstellar medium. However, it is usually assumed that the dust grains are spherical and their optical properties can be described by the Mie theory \citep[see, e.g., ][]{1998asls.book.....B} even though the shapes of astrophysical dust grains are unknown and are certainly not perfect spheres \citep{1995ApJ...444..293K}.

It has become clear that the simplifications made regarding the grain shapes lead to difficulties in interpreting the infrared (IR) spectral energy distributions (SEDs) and spectra of nebulae around evolved stars. For example, in order to model the long wavelength IR SEDs of some PN and WRN, authors have had to postulate extremely high dust-to-gas mass ratios, very large grain sizes or a high degree of porosity in order to reproduce the observations \citep[e.g., ][]{2020MNRAS.497.4128J,Jimenez2021,2021MNRAS.503.1543T}. Furthermore, spherical geometry in dust grains, in comparison with non-spherical grains, implies lower absorption coefficients, leading to lower emission at longer wavelengths \citep{Min2003}.

Several studies have invoked ellipsoidal, rather than spherical, grains to explain the broad 30~$\mu$m feature seen in the infrared spectra of some carbon stars, post-AGB objects and planetary nebulae \citep[e.g., ][] {2002A&A...390..533H,2018A&A...617A..85G} and attributed to magnesium sulphide (MgS, \citealp{1999A&A...345L..39S}). In these applications, the dust grain properties are represented by the continuous distributions of ellipsoids (CDE) formalism described by \citet{1998asls.book.....B}. \citet{2018A&A...617A..85G} find that a mixture of ellipsoidal morphologies is required to reproduce the observations of the 30~$\mu$m feature and the main effect of the ellipsoidal grains compared to spherical ones is to shift the emission to longer wavelengths.

\citet{2021ApJ...910...47D} reviewed the extinction and polarization cross sections of the \citet{1998asls.book.....B} CDEs, together with alternative CDEs that impose constraints on the axial ratios of the ellipsoids in order to avoid the most extreme morphologies and produce more realistic shape distributions \citep{1992A&A...261..567O,1996MNRAS.282.1321Z}. The cross sections can be used with absorption observations to generate self-consistent dielectric functions, which can then be tested against polarization data to determine which CDE best reproduces the observations. However, \citet{2021ApJ...919...65D} caution that the calculation of extinction cross sections at wavelengths comparable to the grain sizes for distributions of ellipsoids is time-consuming, even for a single grain size, and so to do this for a full set of shapes that make up the preferred \citet{1992A&A...261..567O} and \citet{1996MNRAS.282.1321Z} CDEs is impractical. 

In this paper, we adopt an alternative approach based on the database of optical properties of irregular aerosol particles (principally dust and volcanic ash) that has been developed for terrestrial remote-sensing applications by \citet{Saito2021}. The motivation behind this database (Texas A\&M University dust 2020: TAMUdust2020) is the awareness that models of optical properties based on spherical and spheroidal particle shapes are inadequate for retrieving the physical properties of the aerosol particles and lead to uncertainty in the interpretation of observational data. This has important consequences for understanding the global distribution of aerosol properties and their impact on the climate. These are similar concerns to those that arise from astrophysical applications. The applicability of TAMUdust2020 to irregular dust grains has recently been demonstrated by \citet{Martikainen2025}, who used it to model the optical properties of Martian dust analogues. Combining the database with measured particle size distributions and derived complex refractive indices, they achieved good agreement with laboratory scattering data, validating the hexahedral model for realistic dust particles.

The TAMUdust2020 database assumes that the particles are composed of collections of irregular hexahedra. The available information consists of the single-scattering properties of these particles calculated using state-of-the-art computations for the size parameter range between the Rayleigh scattering domain up to 11800, a refractive index range typical of dust and volcanic ash particles, and a degree of sphericity range from 0.695 to 0.785 \citep{Saito2021}. It can be used for radiative transfer applications from the ultraviolet to the infrared, and detailed comparison of the database-calculated values with laboratory measurements from the Granada-Amsterdam Light Scattering Database \citep{Munoz2025} shows good agreement. The ranges covered by TAMUdust2020 overlap with those of typical circumstellar dust and in this paper we propose the first astrophysical application, to our knowledge, of the database. 

This paper is organized as follows: in \S\ref{sec:methods} we describe how we obtain the optical properties of dust grains of astrophysical interest, while \S\ref{sec:results} describes applications to photoionization modelling circumstellar nebulae around evolved stars and makes comparisons with models using spherical grains. In \S\ref{sec:discuss} we discuss the relevance of our models and \S\ref{sec:summary} summarizes our findings. 

\section{Methods}
\label{sec:methods}
\subsection{TAMUdust 2020 database}
\label{ssec:tamu}
We used the TAMUdust2020 \citep{Saito2021} database to obtain the single-scattering properties of silicate, graphite  and amorphous carbon grains. The main assumption of TAMUdust2020 is that the grain shapes are irregular hexahedra, characterized by their maximum diameter $D$ and aspect ratio, which is quantified by the sphericity parameter $\Psi$. 

The particular mathematical method used to calculate ensemble grain-scattering properties in TAMUdust2020 depends on the particle size parameter, defined as
\begin{equation}
    X = \pi \frac{D}{\lambda}  \ ,
\end{equation}
where $\lambda$ is the wavelength. For very small size parameters ($X \ll 1$), the Rayleigh scattering approximation is used. For small to moderate values of the size parameter the invariant imbedding T-matrix approximation (IITM) is used \citep{1988ApOpt..27.4861J}, while for large size parameters the physical geometrical optics method (PGOM) of \citet{1996ApOpt..35.6568Y,1997JOSAA..14.2278Y} is adopted. 

TAMUdust2020 formats data in two domains: the short-wavelength (SW) kernel and the long-wavelength (LW) kernel. The SW kernel covers size parameter values from the Rayleigh scattering domain up to 11800, while the LW kernel goes up to $X=1470$. The two domains in combination cover a large part of the most relevant grain sizes. The ranges of the complex refractive indices in TAMUdust2020 are restricted to $1.37 \le m_\mathrm{r} \le 1.7$ and $0.0001 \le m_\mathrm{i} \le 0.1$ for the SW kernel, and $0.4 \le m_\mathrm{r} \le 3.2$ and $0.001 \le m_\mathrm{i} \le 4.0$ for the LW kernel. The sphericity parameter $\Psi$ ranges from 0.695—0.785.

The user constructs their own single-scattering property database as a subset of the main TAMUdust2020 database by specifying the wavelength bins ($\lambda$) and maximum diameter bins ($D$) of interest, together with the relevant value of the sphericity parameter ($\Psi$). They must also provide the complex refractive indices that correspond to the particular grain species (e.g., silicate). TAMUdust2020 performs a multidimensional linear interpolation in $(m_\mathrm{r},m_\mathrm{i},\Psi)$-space, assuming  randomly oriented dust particles, and the output consists of the single-scattering properties at the user-specified wavelengths and maximum diameters obtained through spline interpolation in $\ln X$-space \citep[see ][for full details]{Saito2021}.

The output data for each user-specified wavelength and maximum diameter value comprises the volume ($V$) of a particle and its projected area ($A$), together with the extinction efficiency ($Q_\mathrm{ext}$), the single-scattering albedo ($\omega$), asymmetry factor ($g$), and the phase matrix elements. In the present work we are not interested in the phase matrix elements.

\subsection{Dust species}

\begin{table*}
\begin{center}
\begin{tabular}{lcccccccccc}
\hline                                     
Bin                      &B$_1$ & B$_2$ &B$_3$ & B$_4$ &B$_5$ & B$_6$ &B$_7$ & B$_8$&  B$_9$ & B$_{10}$    \\ 
\hline
$a_\mathrm{min}$ [$\mu$m]   & 0.00500 & 0.00739 & 0.01093 & 0.01617 & 0.02319 & 0.03536 & 0.05228 & 0.07731 &  0.11433 & 0.16906    \\
$a_\mathrm{max}$ [$\mu$m]   & 0.00739 & 0.01093 & 0.01617 & 0.02319 & 0.03536 & 0.05228 & 0.07731 & 0.11433 &  0.16906 & 0.25000    \\
\hline
\end{tabular}
\caption{Smallest ($a_\mathrm{min}$) and largest ($a_\mathrm{max}$) grain sizes in each bin within the total size distribution.
}
\label{tab:bines}
\end{center}
\end{table*}

Our primary interest is to study non-spherical grains in the context of photoionization modelling of circumstellar nebulae around evolved stars. AGB stars produce copious amounts of dust in their cool atmospheres during the thermal pulse stage. The chemical composition of the dust depends on the surface abundances of the star during this stage and, in particular, on the C/O ratio \citep{2015MNRAS.449.1797D}. If oxygen dominates, then oxygen-rich grains such as oxides and silicates are formed. On the other hand, if carbon dominates, then carbon dust and carbides such as SiC and TiC, and sulphides like MgS  can be expected. Low-mass stars ($<1.5 M_\odot$), which do not undergo surface chemical enrichment in their late evolution, and progenitors with masses above 4 to 5~$M_\odot$, which undergo hot bottom burning that destroys carbon at the base of the convective envelope, have surface abundances dominated by oxygen and will produce silicate dust. Intermediate mass stars, which undergo multiple third dredge-up events during the thermal pulse stage, have surface abundances dominated by carbon. 

\citet{2018A&A...617A..85G} find evidence for large amorphous carbon grains ($a<0.4\mu$m), small graphite grains ($a<0.03\mu$m), SiC and magnesium and iron sulphides in the PN IC 418. \citet{2002A&A...382..222M} used Infrared Space Observatory observations in the 2 to 200~$\mu$m wavelength range to study oxygen-rich dust shells around 17 evolved stars. They found broad emission bands, which can be attributed to amorphous silicate dust, and many narrow band emission features, which they attributed to crystalline silicate dust. Comparison with laboratory data revealed that the narrow-band emission was due to both olivines (Mg$_{2x}$Fe$_{2-2x}$SiO$_4$) and pyroxines (Mg$_{x}$Fe$_{1-x}$SiO$_3$) but that the dust was Mg-rich and Fe-poor, i.e., $x\sim 1$. This general finding has been confirmed in many objects since. \citet{2017MNRAS.466.1963L} and \citet{2013ApJ...765...72J} found that silicate dust in oxygen-rich evolved stars has a high degree of crystallinity, unlike the interstellar medium, which contains predominantly amorphous silicate dust. 

In the case of WRN, two different processes can produce circumstellar dust around Wolf-Rayet (WR) stars, whose progenitors are massive stars with initial masses $M_\mathrm{init} > 25 M_\odot$. The first is associated with ring nebulae, which are comprised principally of swept-up material that was expelled due to mass loss in a previous evolutionary stage of the star. This dust will have formed during the preceding red supergiant (RSG), yellow supergiant (YSG) or luminous blue variable (LBV) stage and observational evidence for amorphous silicates, crystalline silicates, PAHs and metallic iron has been found in this type of object \citep{2010ASPC..425..267W,2014A&A...569A..80G}. The ratio of amorphous to crystalline silicate dust appears to be related to the mass-loss rate, with higher mass-loss rates favoring a higher degree of crystallinity. The other process occurs in eccentric colliding wind binary systems, where the primary is a WC (carbon-rich) WR star and the secondary is an O star \citep[see, e.g.,][]{2025ApJ...987..160R}]. Carbon-rich dust condenses in the dense interaction region at periastron.

The dominant grain species in photoionized nebulae around evolved stars are amorphous carbon and silicates, and to a lesser extent graphite, and so these grain species will be the focus of the present paper.

We take the optical properties for these grain species from the data repository distributed with the \textsc{cloudy} v17.00 photoionization code \citep{Chatzikos2023}, that is, the refractive index of graphites and astronomical silicates come from \cite{MartinRouleau1991}, and for amorphous carbon come from \cite{RouleauMartin1991}. Note that “astronomical silicate” has a synthesized dielectric function based on crystalline olivine but requires that the optical properties of the grains be consistent with several observational constraints, such as absorptivity consistent with circumstellar emission from oxygen-rich stars and emissivity consistent with hot dust in the Trapezium region of the Orion nebula \citep{1984ApJ...285...89D}.

\subsection{Size distribution}

The size distribution of grains in circumstellar nebulae around evolved stars is governed by grain-grain collisions in the grain-forming stellar atmosphere. \citet{1980ApJ...241L.105B} show that the emergent grain radii, $a$, should follow a power law  $N(a) \propto a^{-3.5}$. This agrees with the size distribution of grains in the diffuse interstellar medium, commonly described by the MRN power law \citep{Mathis1977}. The minimum and maximum grain sizes of the size distribution depend on the astrophysical context and are often established through modelling of observed infrarred spectra \citep[see, e.g.,][]{2020MNRAS.497.4128J,Jimenez2021}. 

In this paper, we adopt minimum and maximum grain sizes $a_\mathrm{min}=0.005 \mu$m and $a_\mathrm{max}=0.25 \mu$m, respectively. Moreover, we divide the power-law size distribution into 10 bins with constant ratio  $a_\mathrm{i,max}/a_\mathrm{i,min}=(a_\mathrm{max}/a_\mathrm{min})^{1/10}=1.47876$ (see Table~\ref{tab:bines}). Dividing the full size range into bins enables us to differentiate the contributions of small grains (which absorb most of the radiation) from those of the large grains (which contain most of the mass). 

From TAMUdust2020, using the procedure described in section~\ref{ssec:tamu}, we obtained the single-scattering properties of 140 different values of particle diameters $D$ ranging from 0.001—3.0~$\mu$m at a discrete set of wavelength values in the optical to infrared part of the electromagnetic spectrum, where all particles are assumed to have the same sphericity, $\Psi=0.7$. Note that all of our data corresponds to the LW domain due to the grain size range we considered. In order to avoid resonances that can occur when single grain sizes are used \citep{1998lsnp.conf.....M}, we calculate the bulk scattering parameters of our hexahedral grains by integrating over each size bin, $[D_\mathrm{i,min},D_\mathrm{i,max}]$, of the grain size distribution shown in Table~\ref{tab:bines}, where $D_k=2a_k$. This is accomplished by constructing spline interpolations of each of the TAMUdust2020 output parameters at the bin boundaries and 10 interior points and then evaluating the bulk properties by numerical integration. The equivalent diameter, $D_\mathrm{eq} = 2 \sqrt{A/\pi}$,  where $A$ is the projected area, is used to define the “knot” points of the splines.

The particle size distribution for the power-law assumed in this work is
\begin{equation}
n(D) = \frac{dN}{dD} \ .
\end{equation}
With this definition, and for a fixed value of the sphericity parameter, the bulk extinction efficiency $\langle Q_\mathrm{ext}\rangle$, the bulk single-scattering albedo $\langle \omega \rangle$, the bulk asymmetry factor $\langle g \rangle$, and the bulk projected area are computed as follows \citep[cf. ][]{Saito2021}:

\begin{equation}
    \langle C_\mathrm{ext} \rangle =\int_{D_\mathrm{min}}^{D_\mathrm{max}}n(D) A(D) Q_\mathrm{ext}(D,m_r,m_i)~dD
\end{equation}

\begin{equation}
    \langle \omega \rangle =\frac{\int_{D_\mathrm{min}}^{D_\mathrm{max}}n(D) A(D) Q_\mathrm{scat}(D,m_r,m_i)~dD}{\int_{D_\mathrm{min}}^{D_\mathrm{max}}n(D)A(D)Q_\mathrm{ext}(D,m_r,m_i)~dD}
\end{equation}

\begin{equation}
    \langle g \rangle =\frac{\int_{D_\mathrm{min}}^{D_\mathrm{max}}n(D) A(D) Q_\mathrm{scat}(D,m_r,m_i) g(D,m_r,m_i)~dD}{\int_{D_\mathrm{min}}^{D_\mathrm{max}}n(D)A(D)Q_\mathrm{scat}(D,m_r,m_i)~dD}
\end{equation}

\begin{equation}
    \langle A \rangle = \frac{\int_{D_\mathrm{min}}^{D_\mathrm{max}}n(D) A(D)~dD}{\int_{D_\mathrm{min}}^{D_\mathrm{max}}n(D)~dD}
\end{equation}

\begin{equation}
    \langle V \rangle =\int_{D_\mathrm{min}}^{D_\mathrm{max}}n(D) V(D)~dD
\end{equation}

\noindent where $Q_\mathrm{scat} = \omega ~Q_\mathrm{ext}$.

\subsection{Photoionization models}

The main motivation for our study of non-spherical dust grains is to include them in photoionization models of nebulae around evolved stars, where  previous work has shown that the assumption of spherical grains leads to the requirement for very high dust-to-gas mass ratios in order to reproduce the long wavelength IR emission. To this end, we set up a typical two-shell density structure around a hot, evolved star \citep[cf. ][]{Jimenez2021}. The shells represent different mass-loss episodes: generally the outer shell is more dense, with a higher filling factor, while the inner shell is less dense with a lower filling factor. For the present work, we assume a fixed dust-to-gas mass ratio ($\sim 3 \times 10^{-2}$ for carbonaceous grains and $\sim 5 \times 10^{-2}$ for silicates) 
and very similar uniform gas densities in the two shells, since we are not modelling any particular astrophysical object. 

In order to reveal the effects of the hexahedral grains on the continuum emitted by the dusty nebulae around evolved stars, we employed the photoionization code \textsc{cloudy} v.17 \citep{Ferland2017}. This code simulates the physical conditions in a user-defined distribution of dilute gas and dust illuminated by a radiation source, and computes the resulting transmitted spectrum. It includes all the necessary microphysics for both gas and dust and calculates the thermal and statistical equilibrium solution. The dust physics includes grain heating and cooling processes due to the radiation field, collisions with gas particles, stochastic heating and the photoelectric effect, and the grain temperatures are established by the requirement of thermal balance for each grain type and size \citep[for more details see:][]{Ferland2013,Ferland2017,Chatzikos2023}. The default assumption is that the grains are spherical and that their opacities can be described by the Mie theory \citep{vanHoof2001,vanHoof2004}, however, it is possible to include user-specified grain opacities as long as they conform to the \textsc{cloudy} format (see Section~\ref{ssec:gropac}). This is how the TAMUdust2020 data are used in our models.

Inspired by our previous experience modelling dust emission in WRN,
for our \textsc{cloudy} models in the present paper we chose a radiation source that corresponds to a hot, evolved massive star known as a Wolf-Rayet (WR) star. Stellar atmosphere models for such objects have been precomputed by the \textsc{powr} code \citep{Grafener2002,Hamann2003} and are available in the public grid of models presented in \citet{Todt2015}. WR stars are characterized by their strong, line-driven stellar winds, and so their spectra are deficient in the far and extreme ultraviolet regime compared to black bodies of the same effective temperature. We selected model L\,6-13 for the shape of the stellar spectrum, which is the same as that used by \citet{Jimenez2021} for WR\,40, and adopt a typical WR luminosity (see Table~\ref{tab:input}). The final input parameters for the model are the chemical abundances of the gas, for which we use values estimated by  \citet{Esteban2016} and \citet{Mendez2020} for the WRN RCW\,58 around the star WR\,40. The nebular shells are assumed to have constant density and their principal characteristics are listed in Table~\ref{tab:input}.

Note that the procedure for modelling a PN is identical, except that the central star will correspond to a hot, evolved low-mass star, which will typically be an order of magnitude less luminous.

\section{Results}
\label{sec:results}

\begin{table}
\begin{center}
\begin{tabular}{lcc}
\hline                                     
Parameter                              &Inner shell & Outer shell \\ 
\hline
\hline
Distance [kpc]                         & 3.8     &   3.8  \\
$\log_{10}[L_\star/\mathrm{L_\odot}]$        & 5.91    &   5.91   \\
Inner radius["]                        & 146     &   186     \\
Outer radius["]                        & 186     &   228     \\
\hline
$n_{\mathrm{0}}$\,[cm$^{-3}$]          & 16       & 31    \\  
Filling Factor                         & 0.03     & 0.08  \\
\hline
$D/G^{\mathrm{(a)}}$                                  & - &  $2.94 \times 10^{-2}$ \\
$D/G^{\mathrm{(b)}}$                                  & - &  $4.96 \times 10^{-2}$  \\
\hline
\end{tabular}
\caption{Input parameters of the \textsc{cloudy} models. $^{\mathrm{(a)}}$Carbonaceous grains. $^{\mathrm{(b)}}$Silicate grains.} 
\label{tab:input}
\end{center}
\end{table}

\subsection{Grain opacity} 
\label{ssec:gropac}
From the bulk scattering properties computed by TAMUdust2020 we obtained the absorption ($C_\mathrm{abs}$) and scattering ($C_\mathrm{scat}$) cross sections using the expressions:

\begin{equation}
    C_\mathrm{ext} = Q_\mathrm{ext} C_\mathrm{geo},
\end{equation}

\begin{equation}
    C_\mathrm{scat} = \omega ~C_\mathrm{ext},
\end{equation}

\begin{equation}
    C_\mathrm{abs} = C_\mathrm{ext} \left( 1-\omega\right)
\end{equation}

\noindent where $C_\mathrm{ext}$, $C_\mathrm{geo}$, and $C_\mathrm{abs}$ are the extinction, geometric, and absorption cross sections in the wavelegnth range of the optical data available for silicates, graphites and amorphous grains mentioned in the previous section. In other words, we computed values for the absorption and scattering cross sections ($C_\mathrm{abs}$ and $C_\mathrm{scat}$, respectively), as well as the asymmetry parameter ($g$), over a range of size parameters ($X_{1}$, $X_{2}$), where $X_{1} < X_{2}$, or their corresponding wavelengths ($\lambda_{2}$, $\lambda_{1}$), where $\lambda_{2}< \lambda_{1}$. This range was constrained by the availability of refractive index data for the different grain species and the computational limits of TAMUdust2020.

The photoionization code \textsc{cloudy} requires dust extinction data in a specific format: absorption and scattering cross sections per hydrogen nucleon, the difference between unity and the asymmetry factor ($1-g$), and the inverse attenuation length (in cm$^{-1}$). All of these must be provided as functions of the incident photon energy. The required wavelength range spans from 1.297$\times 10^{-8}$$\mu$m to 2.993$\times 10^{7}$$\mu$m\footnote{This corresponds to an energy range from 3.040$\times 10^{-9}$~Ryd to 7.354$\times 10^{6}$~Ryd, as specified in the \textsc{cloudy} documentation.}.

The inverse attenuation length depends on the grain material type and was therefore obtained directly from the \textsc{cloudy} database. In contrast, the other opacity parameters had to be extrapolated to cover the full energy range required by \textsc{cloudy}. In the following sections, we describe the assumptions made for the extrapolation method and present the resulting nebular continuum generated when this extended dataset was used as input for \textsc{cloudy} for each type of dust mentioned in the previous section.

To obtain the absorption and scattering cross section per H nucleon we consider the following normalization:

\begin{equation}
    C~[\mathrm{cm}^2/\mathrm{H}] = C~[\mathrm{cm^2}] \frac{V \times \rho}{\mathrm{MW}\times A}
\end{equation}

\noindent where $\rho$ is the density of the grain material (g/cm$^3$), and MW is the molecular weight of the grain type. $A = A_\mathrm{eff} A_\mathrm{max}$ is the grain abundance; $A_\mathrm{max}$ and $A_\mathrm{eff}$ are the maximum number density of the grain molecule relative to H that can be formed and the efficiency of dust formation, respectively.  $V$ is the volume (cm$^3$) of a dust grain; however, since we analyze a distribution of grain sizes, we take $\langle V \rangle$ calculated for each bin size. 

\begin{figure*}
\centering
\includegraphics[width=0.33\textwidth]{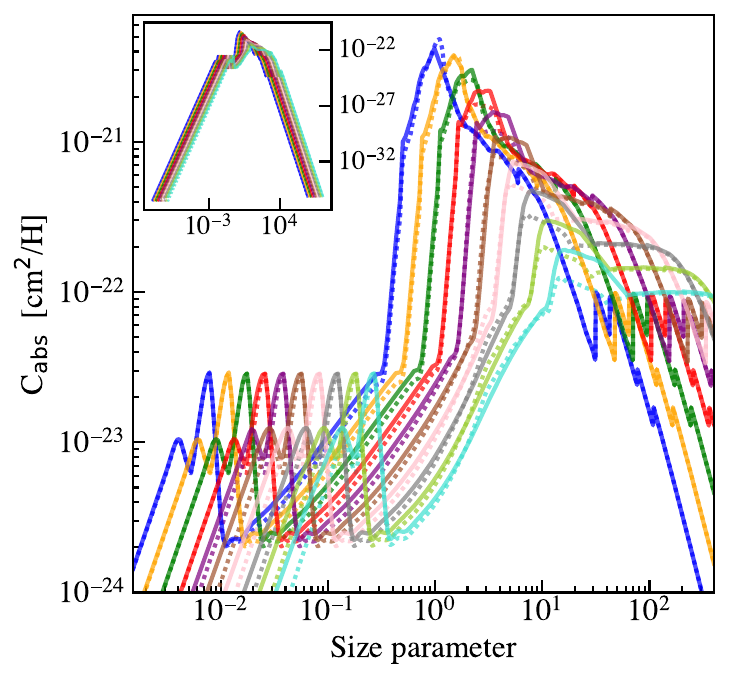}
\includegraphics[width=0.33\textwidth]{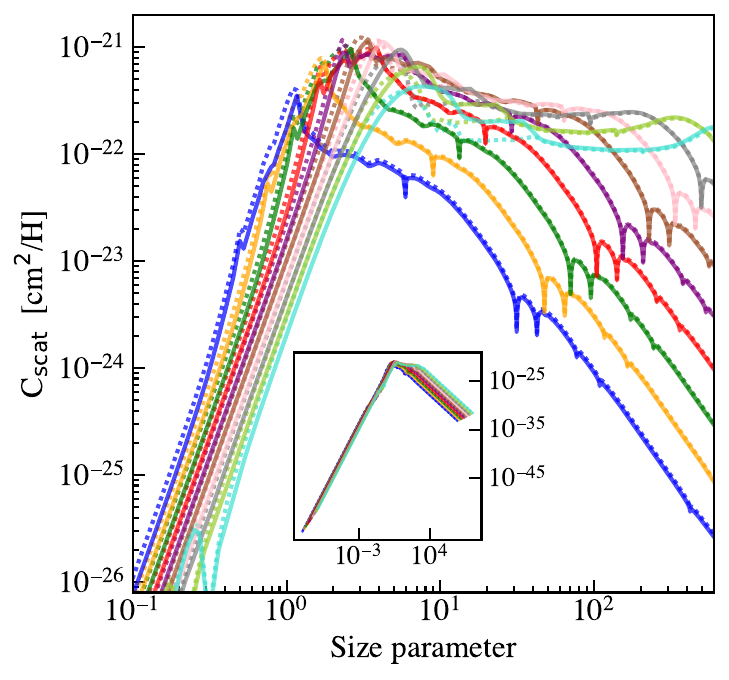}
\includegraphics[width=0.321\textwidth]{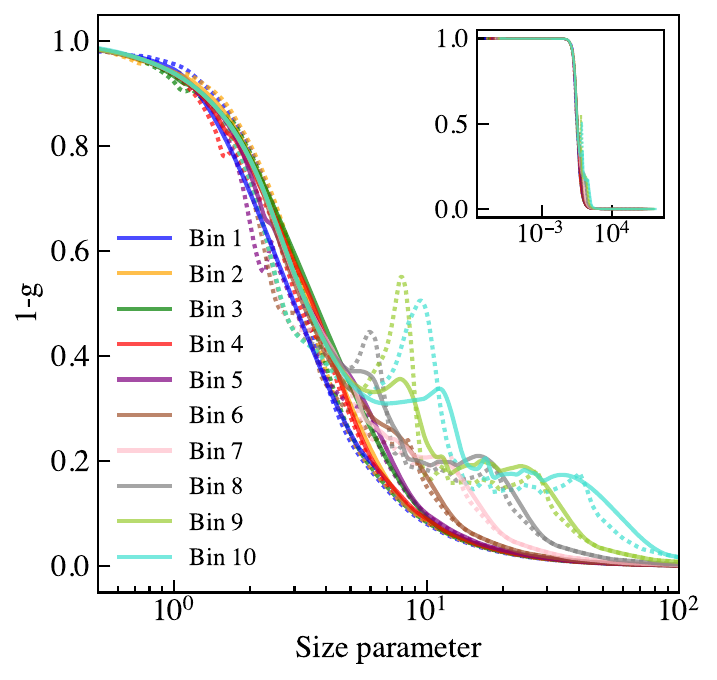}
\caption{Opacity parameters for silicate grains computed for two grain shapes: hexahedral (solid lines) and spherical (dotted lines), across the 10 bins of our size distribution (see Table~\ref{tab:bines}). Left-hand panel: Absorption cross section (C$_\mathrm{abs}$) per H nucleon. Middle panel: Scattering cross section (C$_\mathrm{scat}$) per H nucleon. Right-hand panel: $1-g$ values, where g is the scattering asymmetry factor. Insets display the opacity parameters over the full wavelength range considered in \textsc{cloudy}.}
\label{fig:opacsil}
\end{figure*}

\begin{figure*}
\centering
\includegraphics[width=0.32\textwidth]{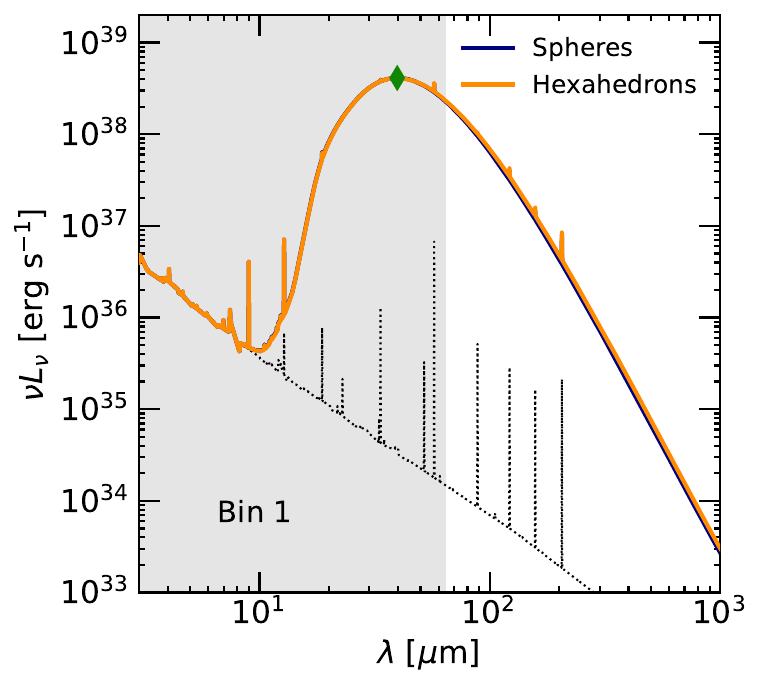}
\includegraphics[width=0.32\textwidth]{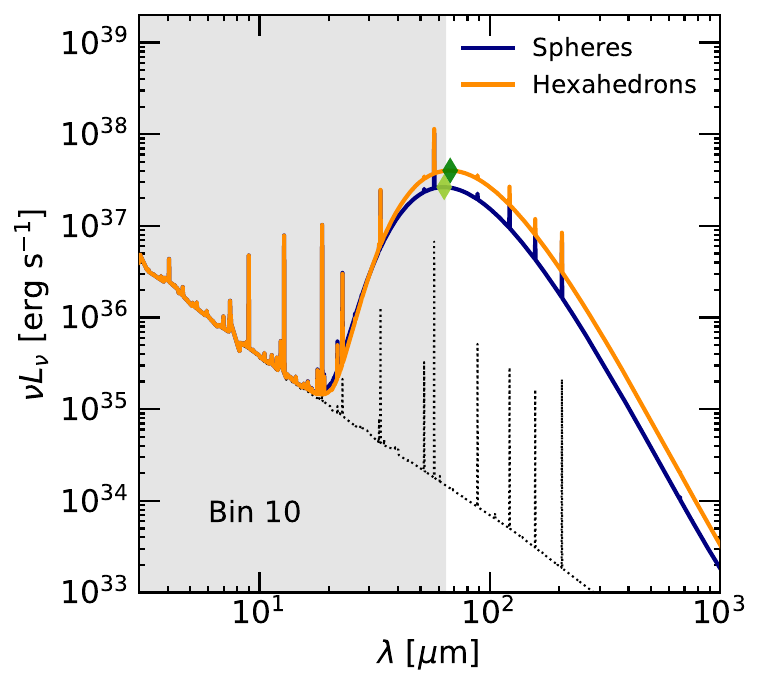}
\includegraphics[width=0.32\textwidth]{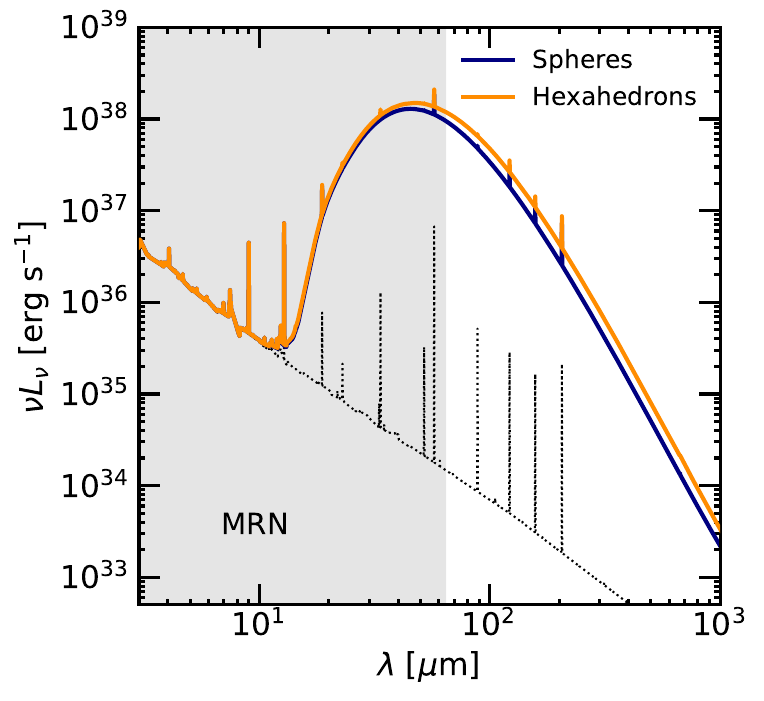}
\caption{Spectra from \textsc{cloudy} photoionization models with silicate grains, comparing spherical (blue line) and hexahedral-shaped (orange line) grains. The first two panels show models using only the grain sizes corresponding to bins 1 and 10 from Table~\ref{tab:bines}. The third panel uses the full MRN size distribution, discretized into the 10 bins listed in the same table. The contribution of the stellar spectrum is shown in black, and green diamonds mark the IR continuum peak of each model. The shaded grey region indicates wavelengths where opacities were derived from TAMUdust2020 output data.}
\label{fig:ModelSil}
\end{figure*}

\subsection{Silicates}

For silicate grains, we computed the absorption cross section ($C_\mathrm{abs}$), the scattering cross section ($C_\mathrm{scat}$), and the scattering asymmetry parameter ($g$) using data from the TAMUdust2020 database over a wavelength range of 0.07~$\mu$m ($\lambda_{\mathrm{sil},2}$) to 64.69~$\mu$m ($\lambda_{\mathrm{sil},1}$). However, the TAMUdust2020 dataset does not fully cover the energy range required by \textsc{cloudy}. To ensure compatibility with the modelling framework, we extrapolated the opacity parameters by considering the grain size parameter, $X$.

\noindent We classified the opacity data into three regimes based on the size parameter:

\begin{itemize}
    \item Set 1: $X<X_{\mathrm{sil},1}$ -- small-particle regime.
    \item Set 2: $X_{\mathrm{sil},1} < X <X_{\mathrm{sil},2}$ -- intermediate regime, based on the opacity values obtained from the TAMUdust2020 database.
    \item Set 3: $X>X_{\mathrm{sil},2}$ -- large-particle regime.
\end{itemize}

In the large-particle regime (\textit{Set 3}), the interaction between light and the surface of the dust particles is more accurately treated via geometric optics. However, due to the computational complexity of modelling this interaction with irregular shapes, we adopted the results for spherical grains\footnote{In this work we refer to opacities calculated for spheres to the opacities calculated by Mie Theory and these are implemented in the base structure of \textsc{cloudy}.} for this regime. Accordingly, for both \textit{Set 1} and \textit{Set 3}, the asymmetry parameter $g$ was taken from spherical grain calculations. 
 
Similarly, for $C_\mathrm{abs}$ and $C_\mathrm{scat}$, to obtain the data in \textit{Set 3} we adopted the values calculated for spheres. In the small-particle regime (\textit{Set 1}), the dipole approximation applies, where opacities vary as $\lambda^{-2}$ and $\lambda^{-4}$ for $C_\mathrm{abs}$ and $C_\mathrm{scat}$, respectively, at long wavelengths \citep[as mentioned by][in the far-IR]{Draine1984}.

Although this approximation is sufficient for the scope of this study, future work could explore efficient methods for extending scattering calculations to irregular grains at large sizes. Identifying convergence trends in the scattering behaviour of hexahedral grains (comparable to those seen in spherical grains) could eliminate the need to sample the entire parameter space currently employed in \textsc{cloudy}.

Figure~\ref{fig:opacsil} displays the values $C_\mathrm{abs}$, $C_\mathrm{scat}$, and $1-g$ as a function of the size parameter, focussing on the data belonging to \textit{Set 2}. For comparison, the values computed for spherical grains are also shown. Insets show the behaviour of the opacity values over the entire wavelength range used in \textsc{cloudy}.  

With full wavelength coverage, we executed a series of \textsc{cloudy} photoionization models. Figure~\ref{fig:ModelSil} and Figure~\ref{fig:ModelSilapp} compare nebular continua produced by models that assume hexahedron-shaped grains versus spherical grains, both of the same size distribution as specified in Table~\ref{tab:bines}.

As shown in Figure~\ref{fig:ModelSil} (see also Figure~\ref{fig:ModelSilapp}), differences in nebular spectra between the two grain shapes are negligible for the smallest grains. However, as the grain size increases, the difference between the continuums becomes more pronounced. This is particularly evident in the IR peak of the nebular spectrum. For models with a dust population of grains with sizes in Bin\,1, the peak wavelengths for spheres and hexahedrons are 39.75~$\mu$m and 39.35~$\mu$m, respectively. For grains with sizes in Bin\,10, the difference is more substantial, with peaks at 63.38~$\mu$m and 67.29~$\mu$m. Corresponding peak luminosities ($\nu L_{\nu}$) are approximately 3.8\% and 52.1\% higher for hexahedral grain models in Bin\,1 and Bin\,10, respectively.

\begin{figure*}
\centering
\includegraphics[width=0.33\textwidth]{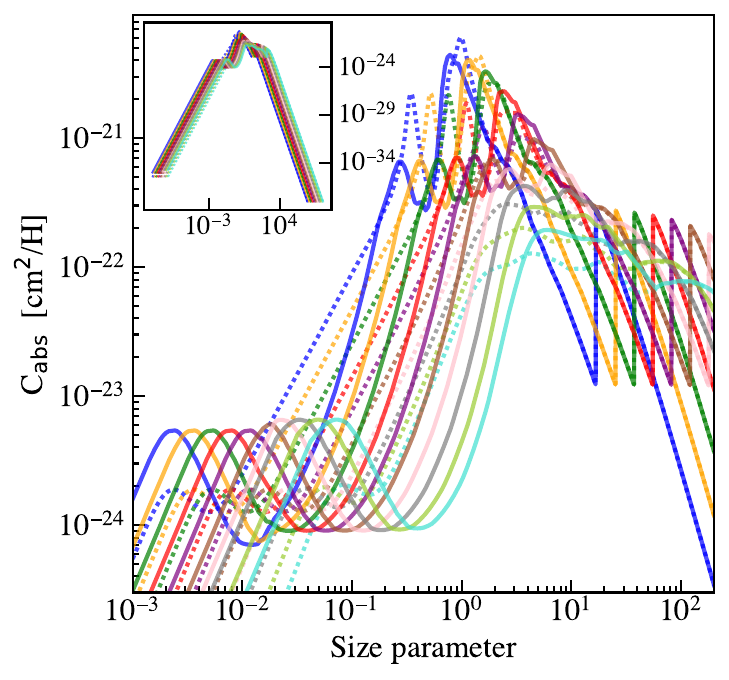}
\includegraphics[width=0.33\textwidth]{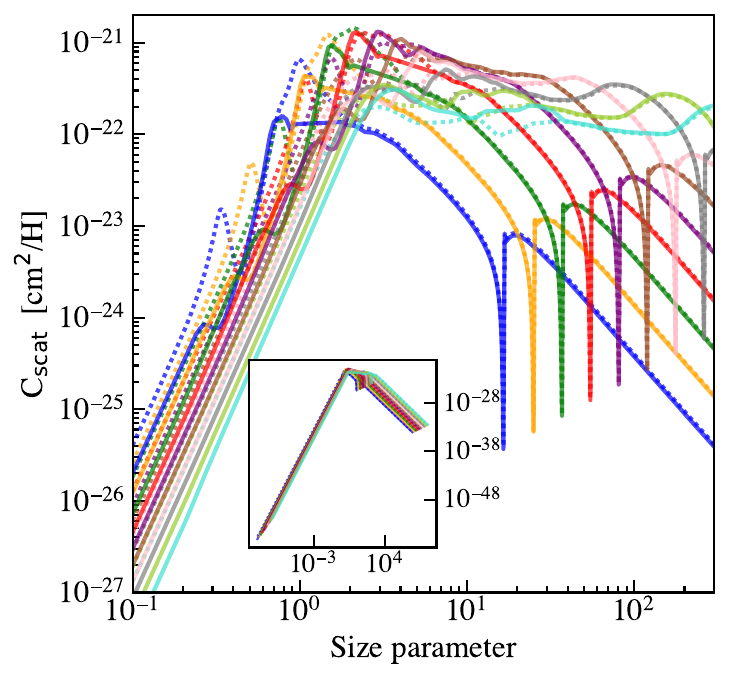}
\includegraphics[width=0.317\textwidth]{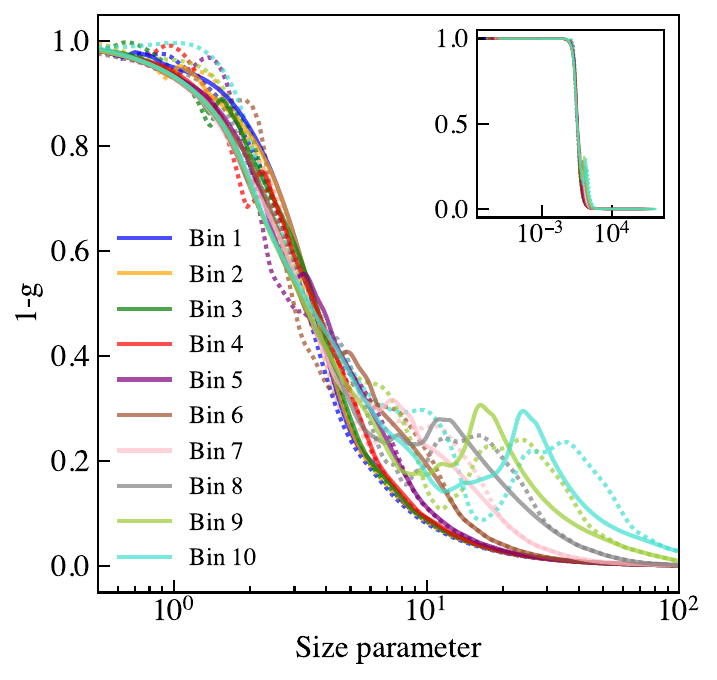}
\caption{Opacity parameters for graphite grains computed for two grain shapes: hexahedral (solid lines) and spherical (dotted lines), across the 10 bins of our size distribution (see Table~\ref{tab:bines}). Left-hand panel: Absorption cross section (C$_\mathrm{abs}$) per H nucleon. Middle panel: Scattering cross section (C$_\mathrm{scat}$) per H nucleon. Right-hand panel: $1-g$ values, where g is the asymmetry factor. Insets display the opacity parameters over the full wavelength range considered in \textsc{cloudy}.}
\label{fig:opacgrap}
\end{figure*}

\begin{figure*}
\centering
\includegraphics[width=0.32\textwidth]{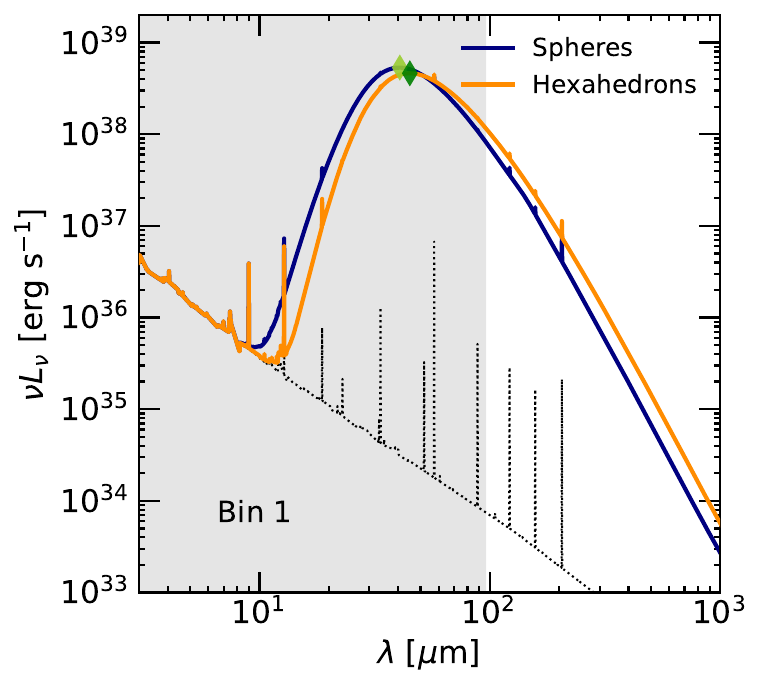}
\includegraphics[width=0.32\textwidth]{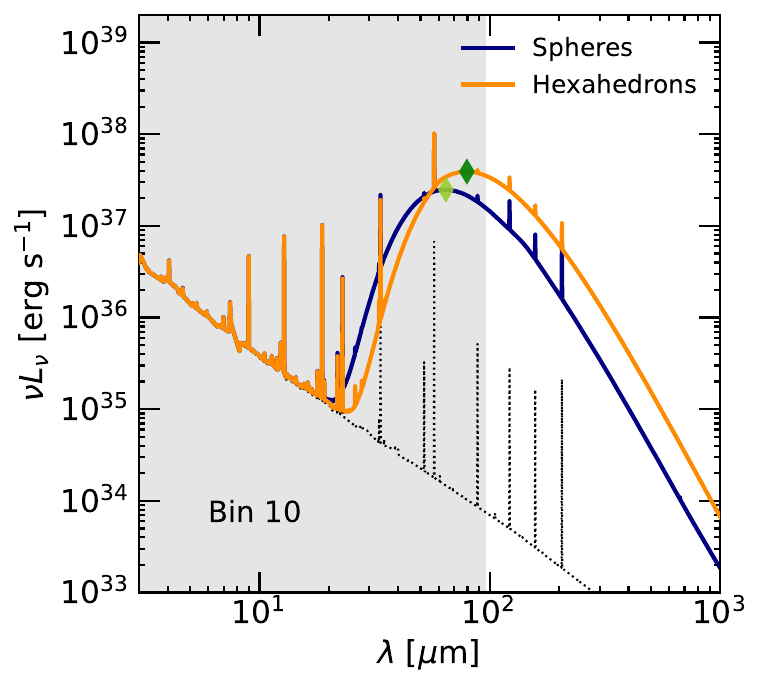}
\includegraphics[width=0.32\textwidth]{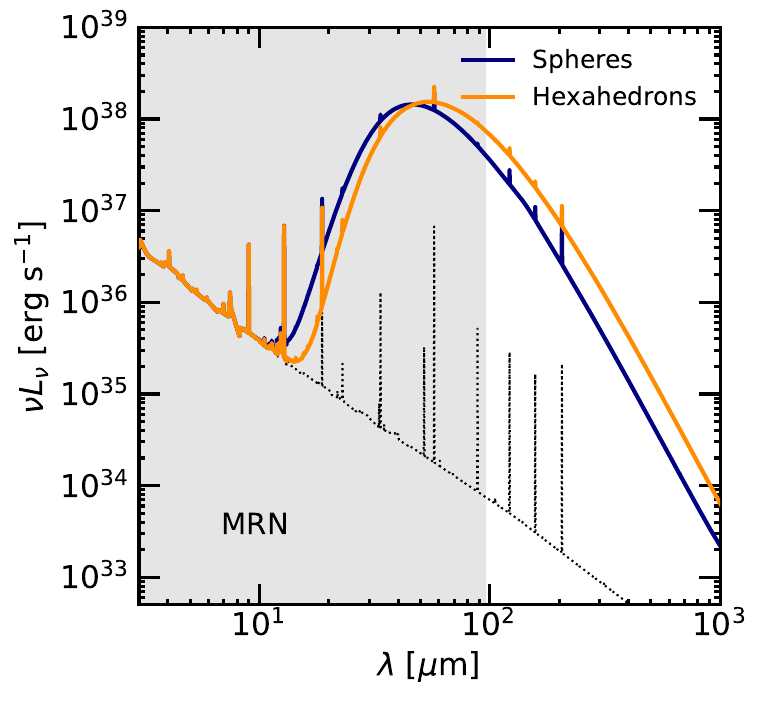}
\caption{Spectra from \textsc{cloudy} photoionization models with graphite grains, comparing spherical (blue line) and hexahedral-shaped (orange line) grains. The first two panels show models using only the grain sizes corresponding to bins 1 and 10 from Table~\ref{tab:bines}. The third panel uses the full MRN size distribution, discretized into the 10 bins listed in the same table. The contribution of the stellar spectrum is shown in black, and green diamonds mark the IR continuum peak of each model. The shaded grey region indicates wavelengths where opacities were derived from TAMUdust2020 output data.}
\label{fig:ModelGrap}
\end{figure*}

\subsection{Graphites}

In the case of graphite grains, we similarly derived $C_\mathrm{abs}$, $C_\mathrm{scat}$, and 1-$g$ from the TAMUdust2020 database over a wavelength range of 0.08~$\mu$m ($\lambda_{\mathrm{gra},2}$) -- 96.33~$\mu$m ($\lambda_{\mathrm{gra},1}$). As with silicates, the data were segmented into three sets according to size parameter, and extrapolated to extend the opacity parameters in the wavelength range required by \textsc{cloudy}. 

Figure~\ref{fig:opacgrap} shows the values of $C_\mathrm{abs}$, $C_\mathrm{scat}$ and $1-g$ as a function of the size parameter, focussing on \textit{Set 2} (obtained from the TAMUdust2020 database). Insets show the behavior of these opacity values over the full wavelength range necessary to run our photoionized models.

Figure~\ref{fig:ModelGrap} and Figure~\ref{fig:ModelGrapapp} show the resulting continuum spectra from the \textsc{cloudy} models assuming hexahedral-shaped and spherical grain geometries. For the smallest grains (Bin 1), the IR peak of the nebular continuum is at  40.68~$\mu$m for spherical grains and at 44.51~$\mu$m for hexahedral grains. For the largest grains (Bin 10), the corresponding IR peaks are located at 64.44~$\mu$m and 79.23~$\mu$m, respectively. Additionally, we obtain a difference in peak luminosity for the smallest grains (Bin 1) of 7.4\% between the two continuums, with the spherical grains yielding the highest value. However, at the IR peak for the largest grains (Bin 10), the hexahedral grains exhibit a luminosity that is 59.7\% higher than that of the spherical grain models.

\subsection{Amorphous carbon}
\subsubsection{AC}
For AC amorphous carbon grains \citep{1987A&AS...70..257B}, we compute $C_\mathrm{abs}$, $C_\mathrm{scat}$, and $1-g$ from TAMUdust2020 output data spanning the range 0.07~$\mu$m ($\lambda_{\mathrm{ac},2}$) -- 239.31~$\mu$m ($\lambda_{\mathrm{ac},1}$). As with previous species, we established \textit{Set 1}, \textit{Set 2}, and \textit{Set 3} according to the value of the size parameter, with \textit{Set 2} covering the broadest range among all species considered in this work.

The asymmetry parameter ($g$), in \textit{Set 1} and \textit{Set 3}, along with $C_{\mathrm{abs}}$ and $C_\mathrm{scat}$ in \textit{Set 3}, were adopted from spherical results. For \textit{Set 1}, $C_{\mathrm{abs}}$ was modeled as a power law, $\propto \lambda^\alpha$, with power index $\alpha \sim 1$ (0.99), consistent with the treatment in \textsc{cloudy v.17} \citep[see][]{Ferland2017}. For $C_\mathrm{scat}$ in \textit{Set 1}, a power law with power index $\sim 4$ (3.86) was used. 

\begin{figure*}
\centering
\includegraphics[width=0.33\textwidth]{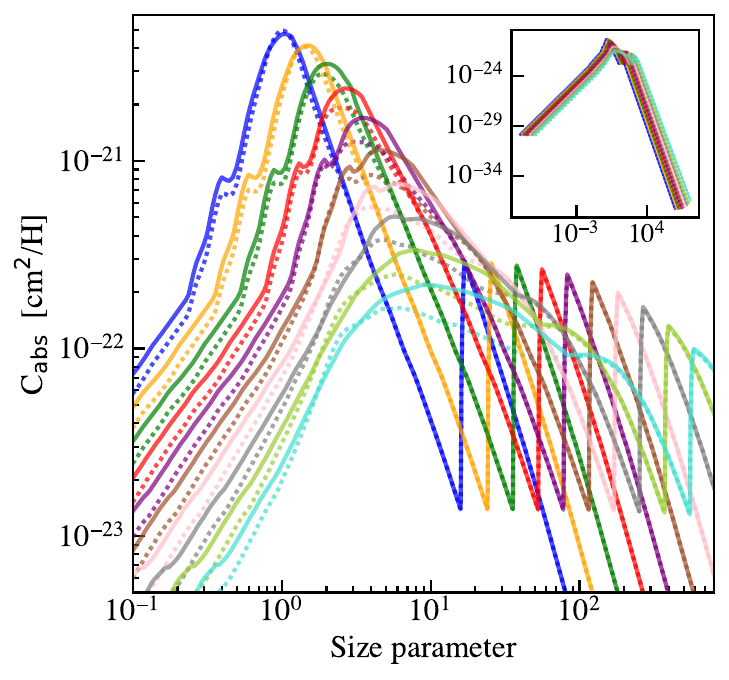}
\includegraphics[width=0.33\textwidth]{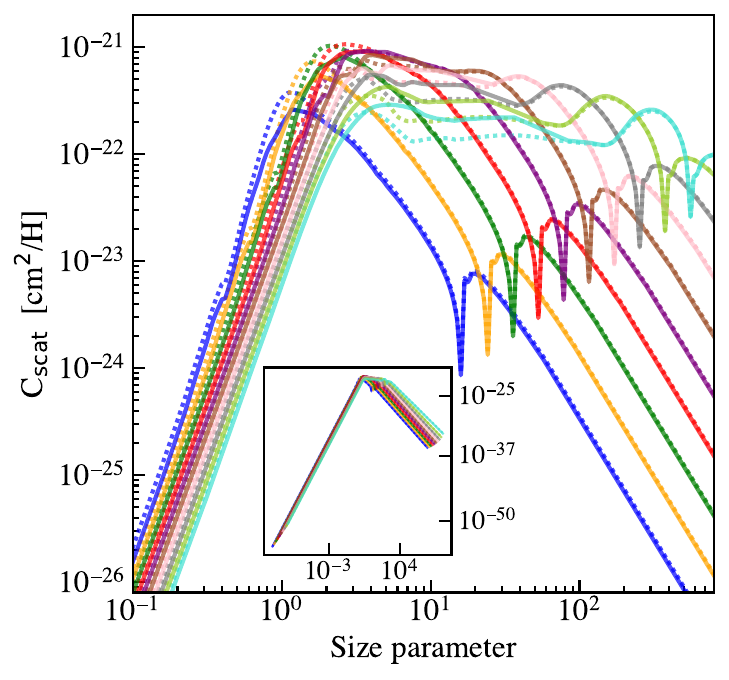}
\includegraphics[width=0.317\textwidth]{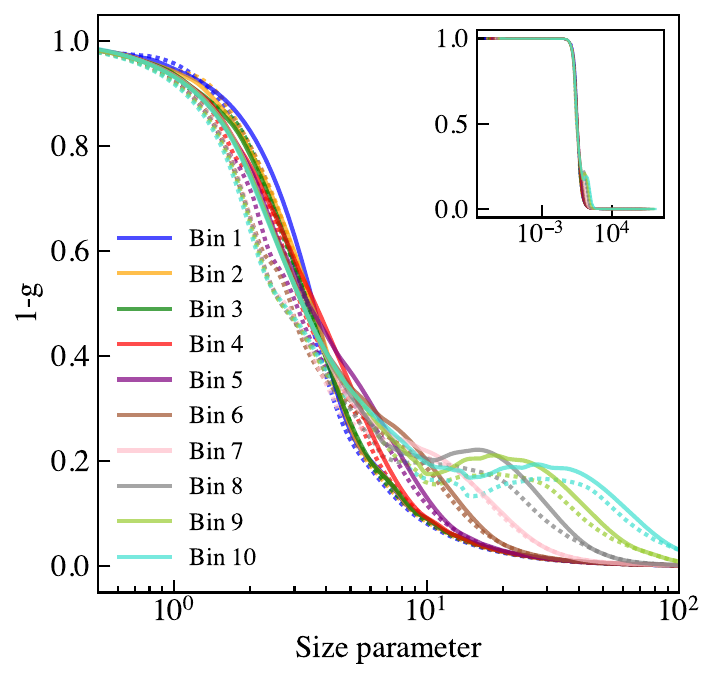}
\caption{Opacity parameters for AC amorphous carbon for two grain shapes: hexahedral (solid lines) and spherical (dotted lines), across the 10 bins of our size distribution (see Table~\ref{tab:bines}). Left-hand panel: Absorption cross section (C$_\mathrm{abs}$) per H nucleon. Middle panel: Scattering cross section (C$_\mathrm{scat}$) per H nucleon. Right-hand panel: $1-g$ values, where g is the asymmetry factor. Insets display the opacity parameters over the full wavelength range considered in \textsc{Cloudy}.}
\label{fig:opacab1}
\end{figure*}

Figure~\ref{fig:opacab1} shows the values of $C_\mathrm{abs}$, $C_\mathrm{scat}$, and $1-g$ as a function of the size parameter, focusing on \textit{Set 2} with insets covering the entire range of the data. 

\begin{figure*}
\centering
\includegraphics[width=0.32\textwidth]{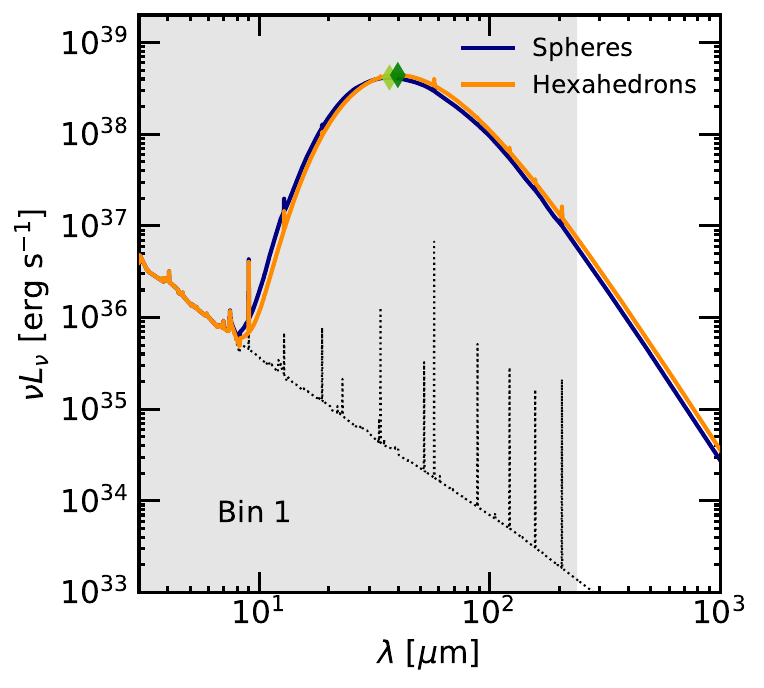}
\includegraphics[width=0.32\textwidth]{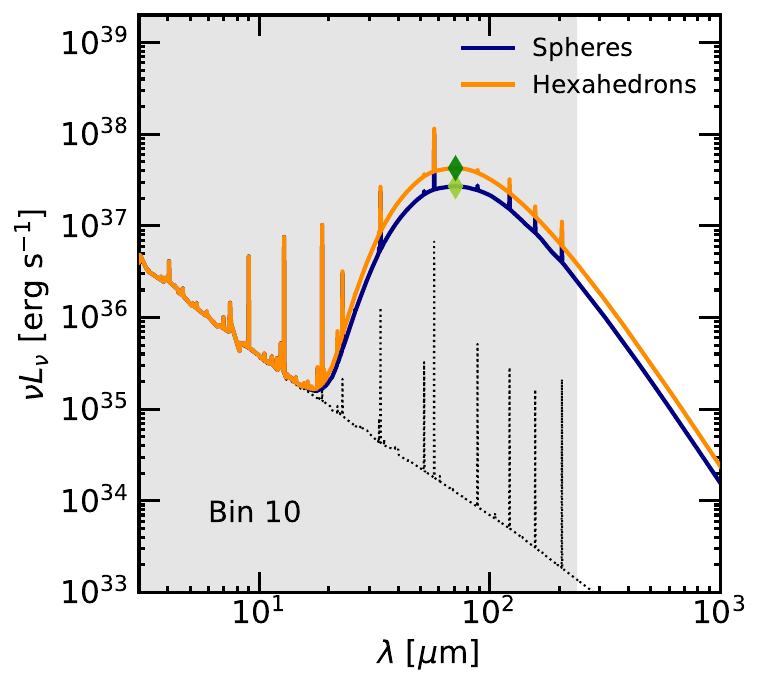}
\includegraphics[width=0.32\textwidth]{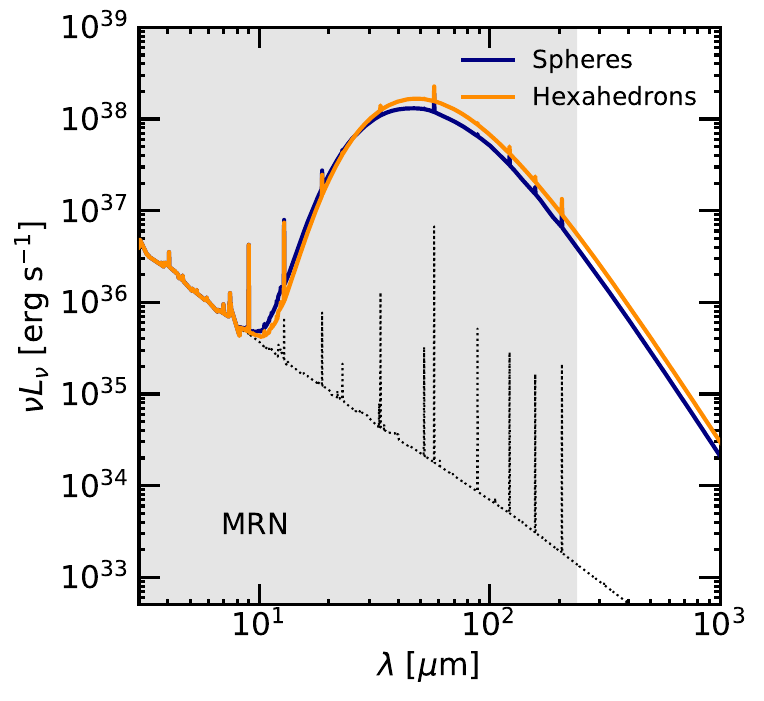}
\caption{Spectra from \textsc{cloudy} photoionization models with AC amorphous carbon grains, comparing spherical (blue line) and hexahedral-shaped (orange line) grains. The first two panels show models using only the grain sizes corresponding to bins 1 and 10 from Table~\ref{tab:bines}. The third panel uses the full MRN size distribution, discretized into the 10 bins listed in the same table. The contribution of the stellar spectrum is shown in black, and green diamonds mark the IR continuum peak of each model. The shaded grey region indicates wavelengths where opacities were derived from TAMUdust2020 output data.}
\label{fig:Modelac1}
\end{figure*}

Figure~\ref{fig:Modelac1} and Figure~\ref{fig:Modelac1app} present the spectra of the \textsc{cloudy} models. For the models with the smallest dust grains, the IR peak wavelengths are at 36.69~$\mu$m and 39.61~$\mu$m for spherical and hexahedral-shaped grains, respectively. For the largest grains, the IR peak wavelengths are at 70.98~$\mu$m for both geometries. The corresponding peak luminosities for the models with hexahedral grains are 6.2\% and 56.2\% higher than the models with spherical grains.

\begin{figure*}
\centering
\includegraphics[width=0.33\textwidth]{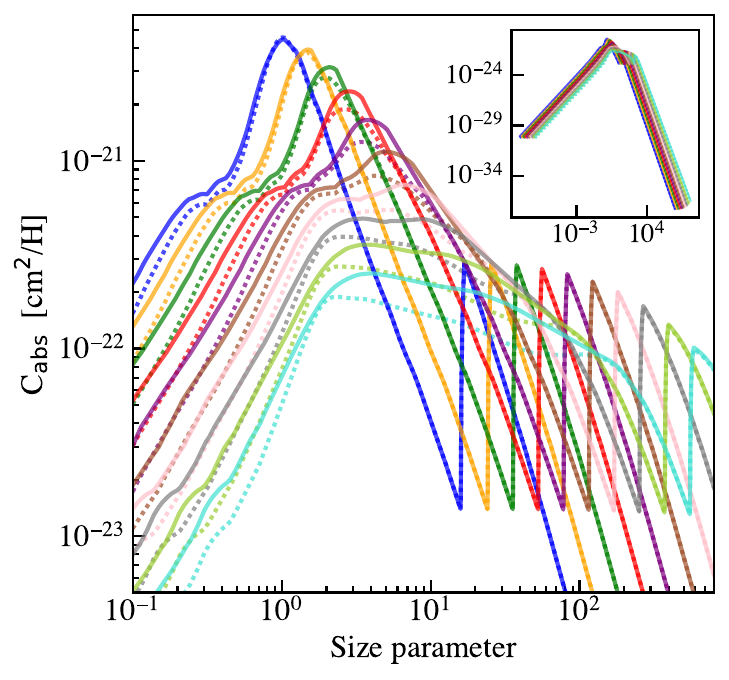}
\includegraphics[width=0.33\textwidth]{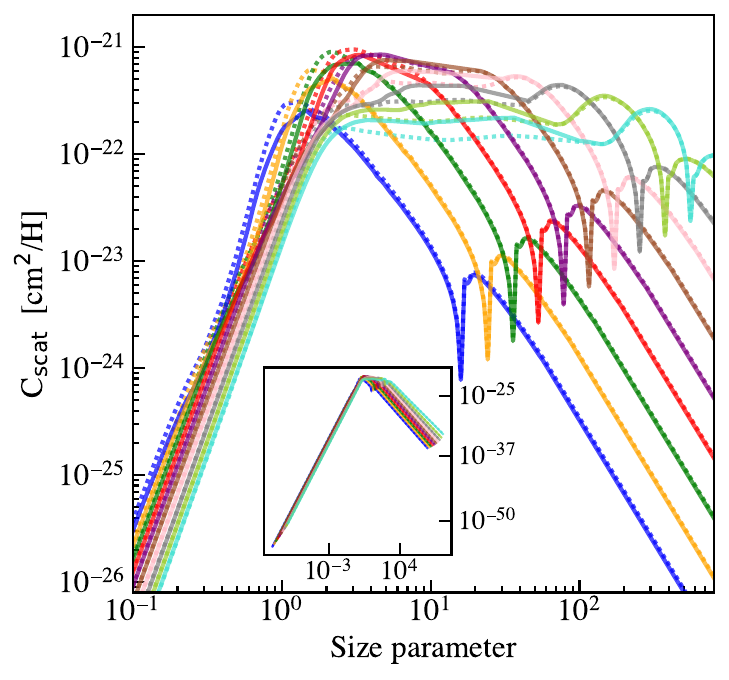}
\includegraphics[width=0.317\textwidth]{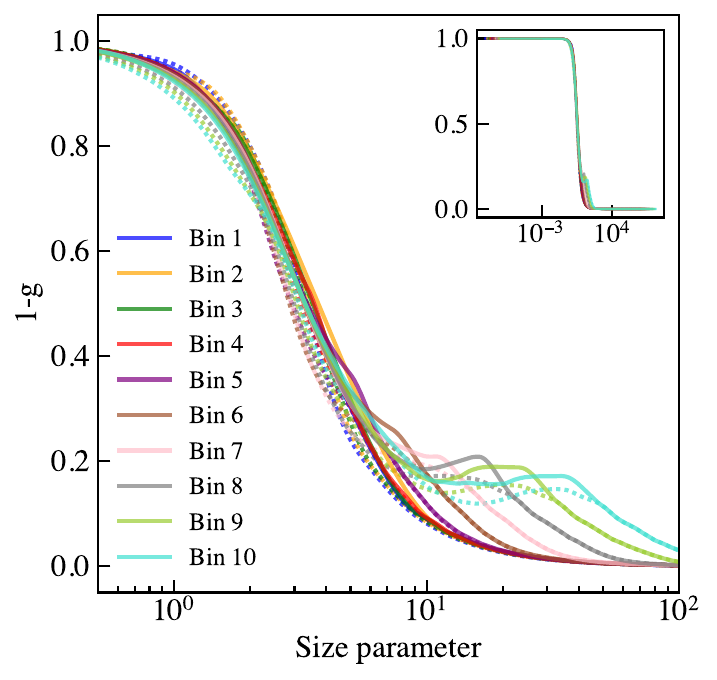}
\caption{Opacity parameters for BE amorphous carbon for two grain shapes: hexahedral (solid lines) and spherical (dotted lines), across the 10 bins of our size distribution (see Table~\ref{tab:bines}). Left-hand panel: Absorption cross section (C$_\mathrm{abs}$) per H nucleon. Middle panel: Scattering cross section (C$_\mathrm{scat}$) per H nucleon. Right-hand panel: $1-g$ values, where g is the asymmetry factor. Insets display the opacity parameters over the full wavelength range considered in \textsc{cloudy}.}
\label{fig:opacbe1}
\end{figure*}

\begin{figure*}
\centering
\includegraphics[width=0.32\textwidth]{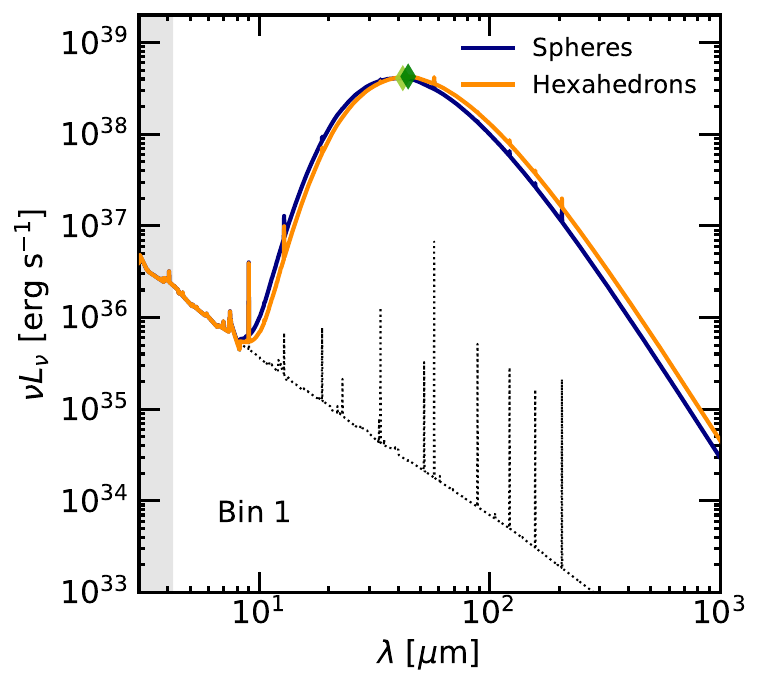}
\includegraphics[width=0.32\textwidth]{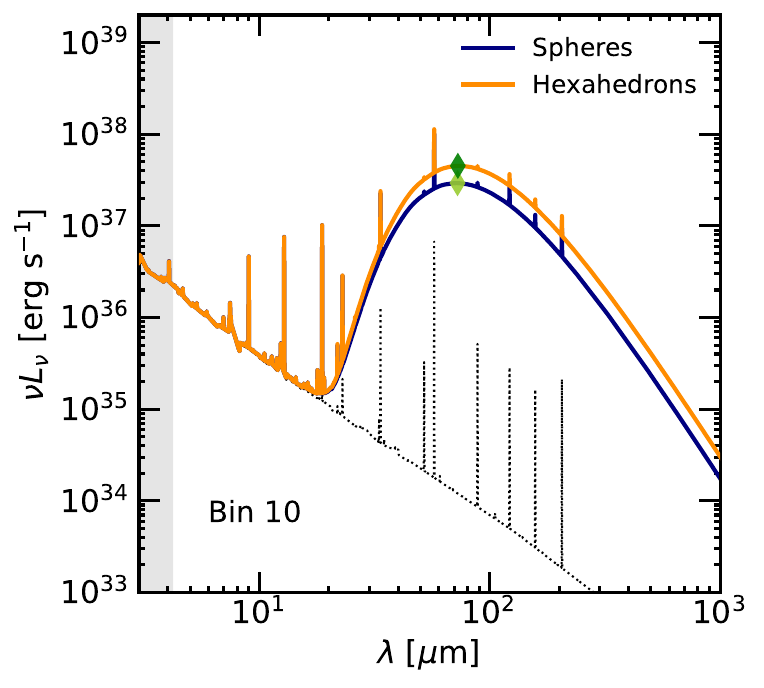}
\includegraphics[width=0.32\textwidth]{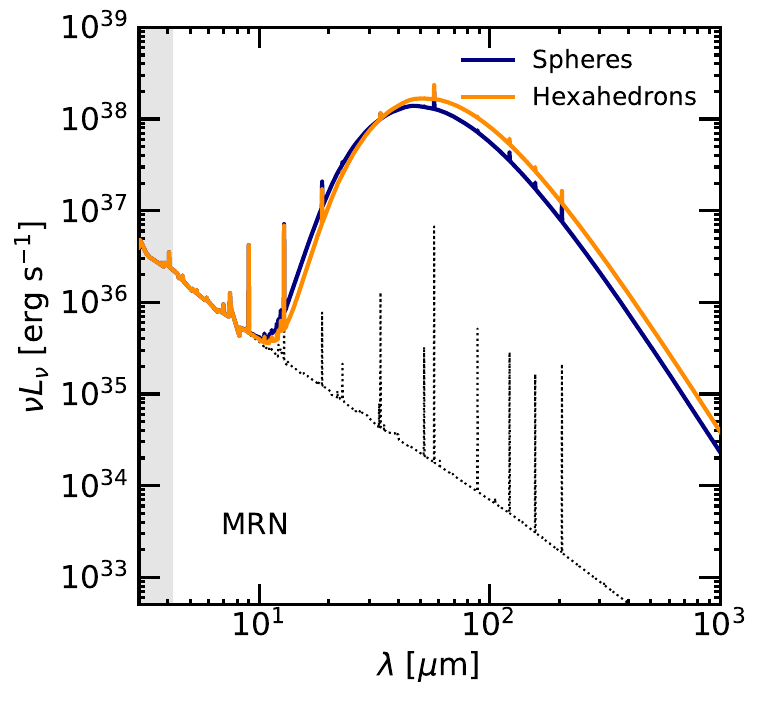}
\caption{Spectra from \textsc{cloudy} photoionization models with BE amorphous carbon grains, comparing spherical (blue line) and hexahedral-shaped (orange line) grains. The first two panels show models using only the grain sizes corresponding to bins 1 and 10 from Table~\ref{tab:bines}. The third panel uses the full MRN size distribution, discretized into the 10 bins listed in the same table. The contribution of the stellar spectrum is shown in black, and green diamonds mark the IR continuum peak of each model. The shaded grey region indicates wavelengths where opacities were derived from TAMUdust2020 output data.}
\label{fig:Modelbe1}
\end{figure*}

\subsubsection{BE}

In the case of BE amorphous carbon grains \citep{1987A&AS...70..257B}, we computed $C_\mathrm{abs}$, $C_\mathrm{scat}$, and $1-g$ from TAMUdust2020 data over the range 0.07~$\mu$m ($\lambda_{\mathrm{be},2}$) -- 2.39~$\mu$m ($\lambda_{\mathrm{be},1}$). As with previous grain species, we worked with the data sets according to the value of the size parameter (\textit{Set 1}, \textit{Set 2} and \textit{Set 3}), with the \textit{Set 2} covering the smallest range among all species considered in this work. 

Once again, we obtain the $1-g$ values for \textit{Set 1} and \textit{Set 3} from asymmetry parameter values estimated from spherical grains approximation.
Similarly, \textit{Set 3} values for $C_\mathrm{abs}$ and $C_\mathrm{scat}$ were adopted from spheres. In the case of \textit{Set 1}, $C_\mathrm{abs}$, and $C_\mathrm{scat}$, follow a power law as $\lambda^\alpha$, with a power law index $\alpha \sim 1$ (0.99) and $\alpha \sim 4$ (3.89), respectively.

Figure~\ref{fig:opacbe1} shows the the values of $C_\mathrm{abs}$, $C_\mathrm{scat}$ and $1-g$ in the region were is \textit{Set 2}. Insets show the values of the opacity parameters in the full range  of the size parameter.

Despite the limited extent of \textit{Set 2}, we were able to run a set of \textsc{cloudy} models. Figure~\ref{fig:Modelbe1} and Figure~\ref{fig:Modelbe1app} show the continuum of these models. For the models with the smallest dust grains, the IR peak wavelengths are at 41.92~$\mu$m and 44.22~$\mu$m for spherical and hexahedral-shaped grains, respectively. Correspondingly, for the largest grains, the IR peak wavelengths are at 72.42~$\mu$m and 72.66~$\mu$m for spheres and hexahedrons, respectively. The corresponding peak luminosities for the models with hexahedral grains are 4.1\% and 54.4\% higher than the models with spherical grains.

\section{Discussion}
\label{sec:discuss}

The models presented in the previous section focus on nebulae around massive evolved stars, specifically Wolf-Rayet stars. However, the primary aim of this work is to establish a starting point for exploring the effects of including irregular dust grains in the modelling of nebulae around evolved stars (with low-, intermediate-, or high-mass central stars) and to provide an initial analysis of the dust continuum emission in these environments.

This approach to including irregular dust grains represents a first step and will require further refinement in future work. Several important aspects merit attention: the inherent complexity of modelling across a broad wavelength range, essential for a self-consistent treatment of both the nebula and its central stellar source; the implications of our findings for the broader study of dust in evolved star environments; and the potential observational consequences of non-spherical grains, particularly in relation to polarization. These topics are discussed in the following sections.

\subsection{Opacity across a broad wavelength range}

\begin{figure*}
\centering
\includegraphics[width=0.24\textwidth]{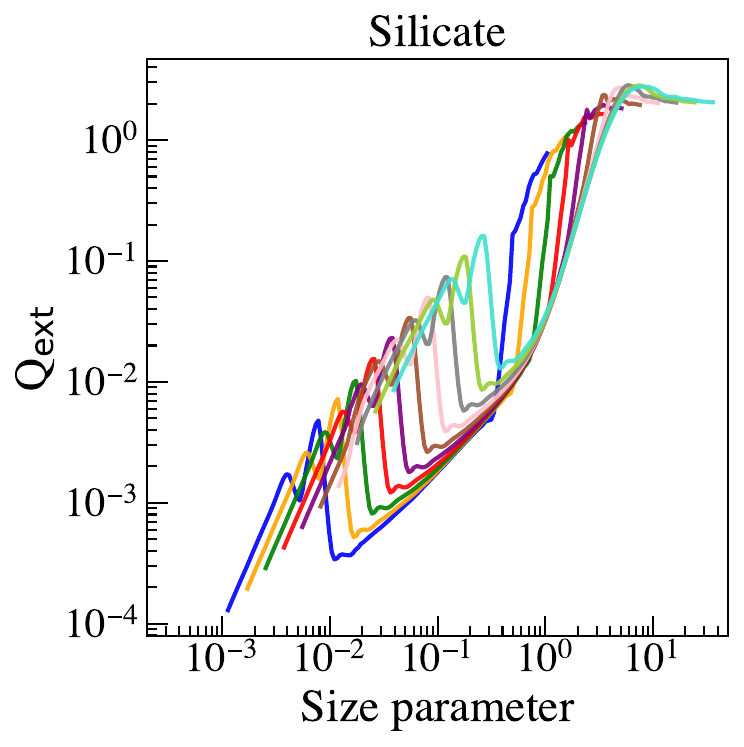}
\includegraphics[width=0.24\textwidth]{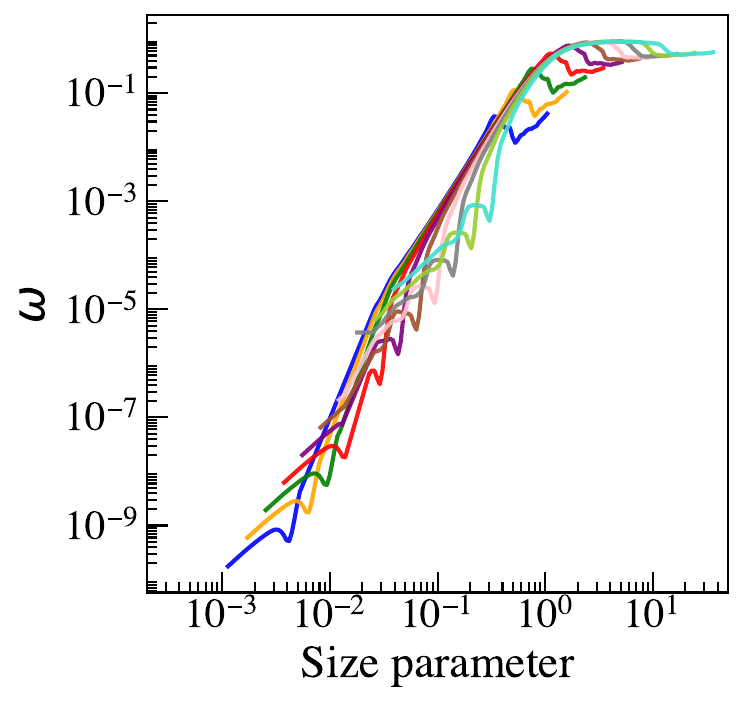}
\includegraphics[width=0.24\textwidth]{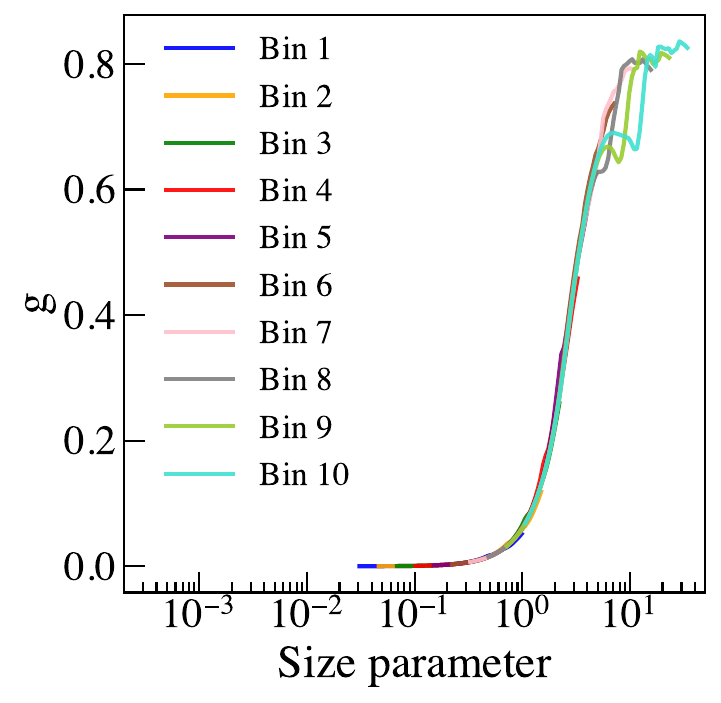}
\includegraphics[width=0.24\textwidth]{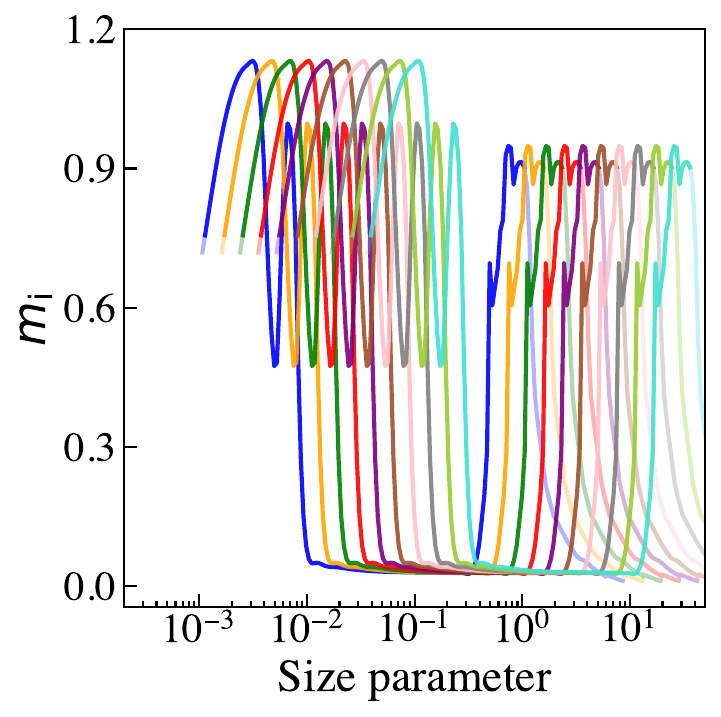}\\
\includegraphics[width=0.24\textwidth]{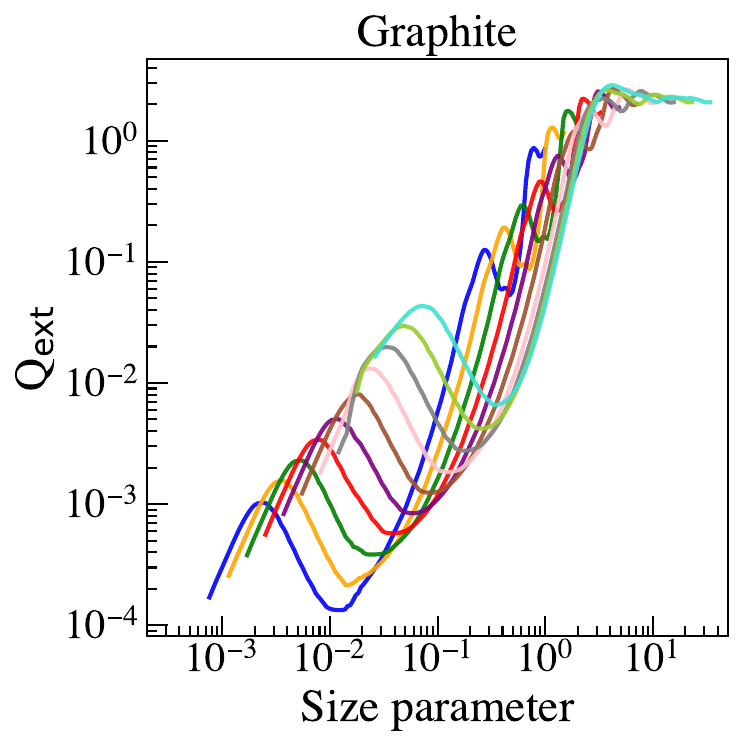}
\includegraphics[width=0.24\textwidth]{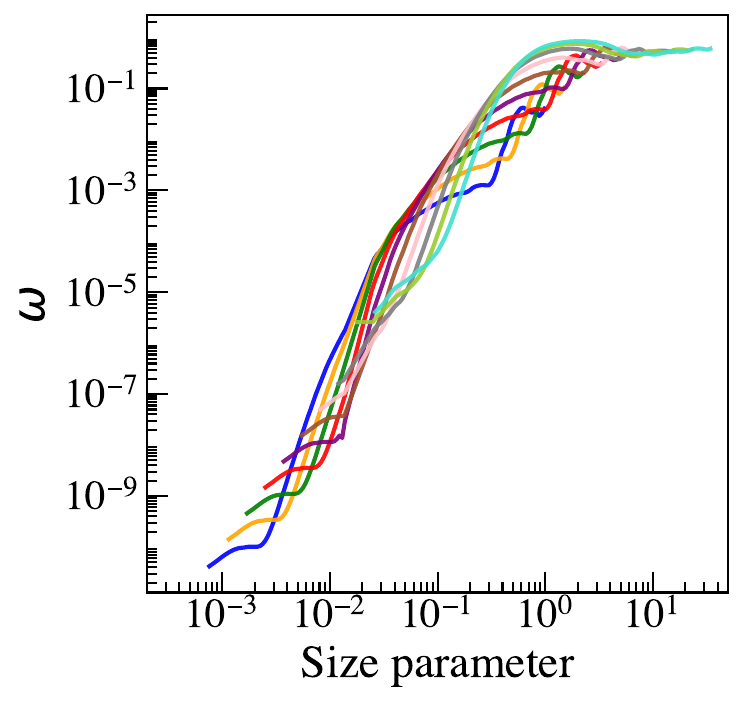}
\includegraphics[width=0.24\textwidth]{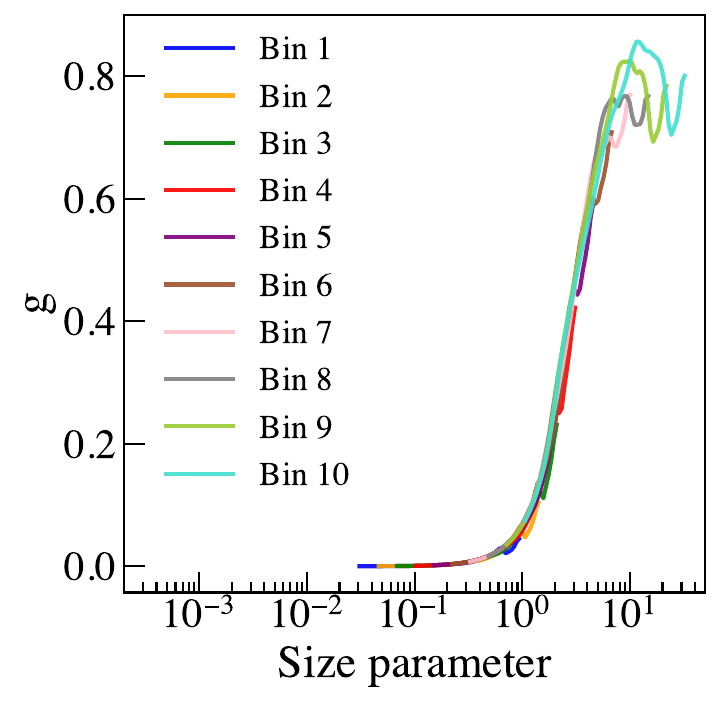}
\includegraphics[width=0.24\textwidth]{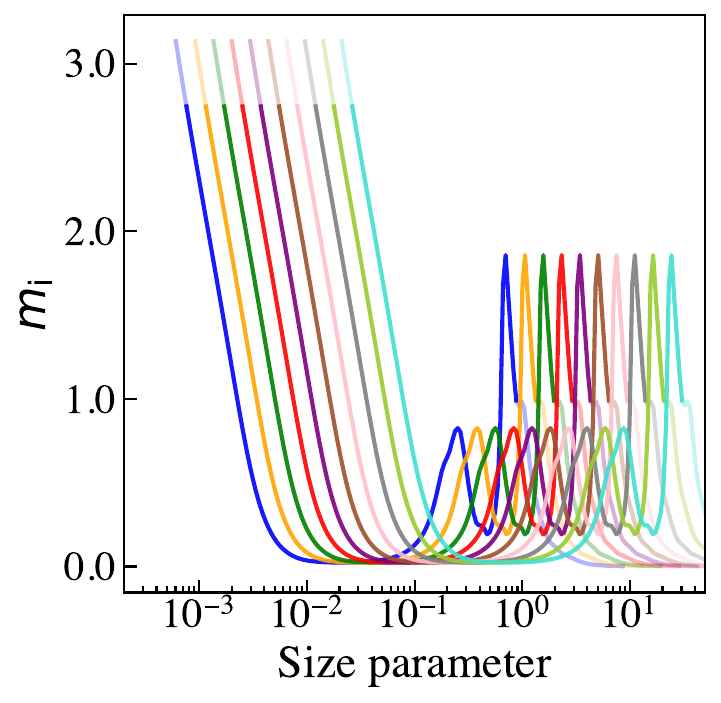}\\
\includegraphics[width=0.24\textwidth]{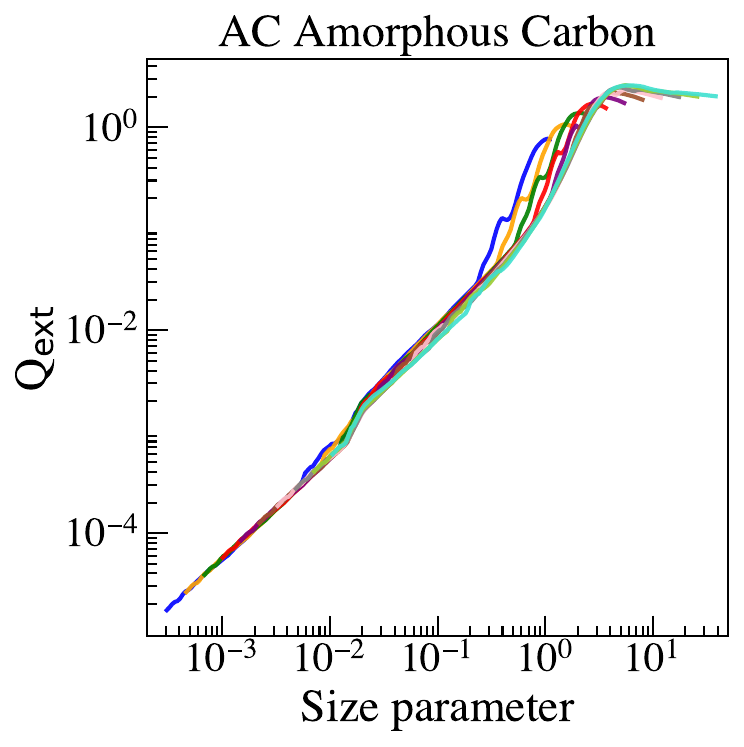}
\includegraphics[width=0.24\textwidth]{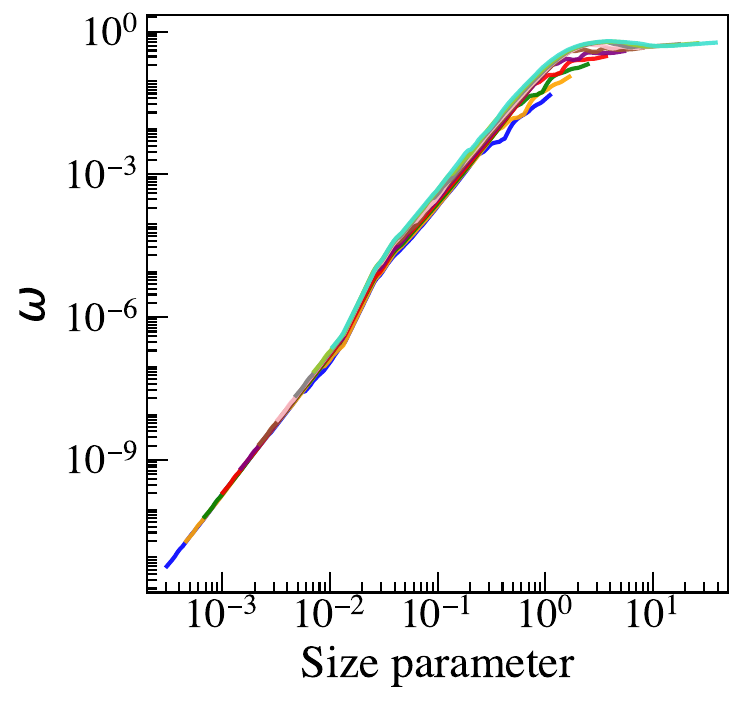}
\includegraphics[width=0.24\textwidth]{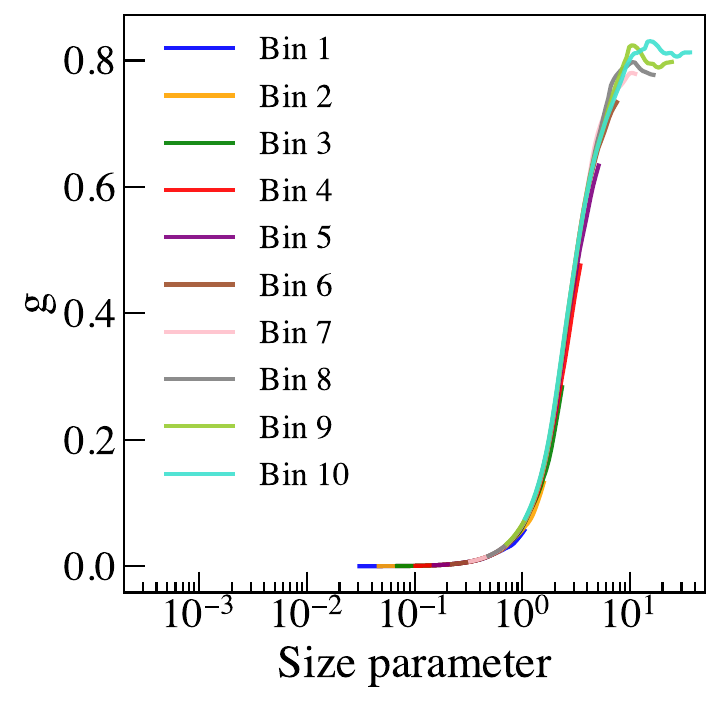}
\includegraphics[width=0.24\textwidth]{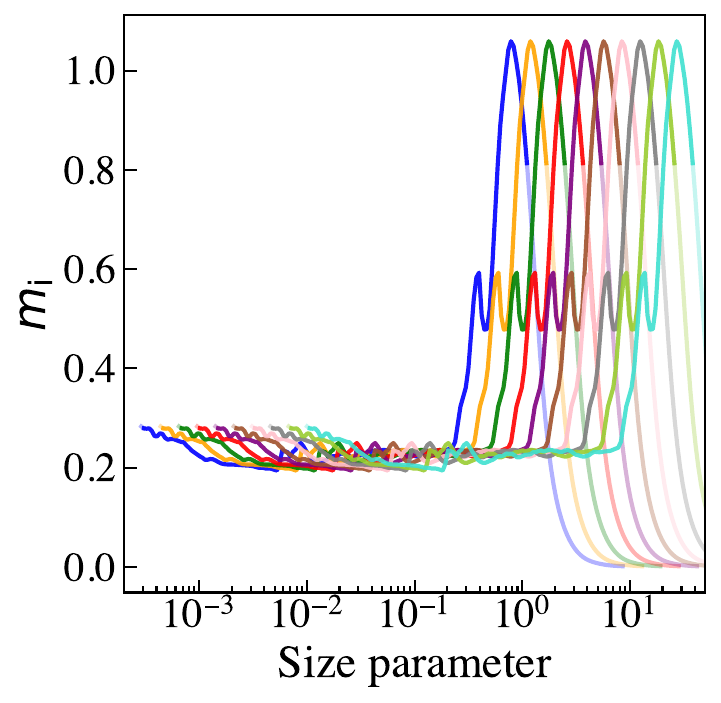}\\
\includegraphics[width=0.24\textwidth]{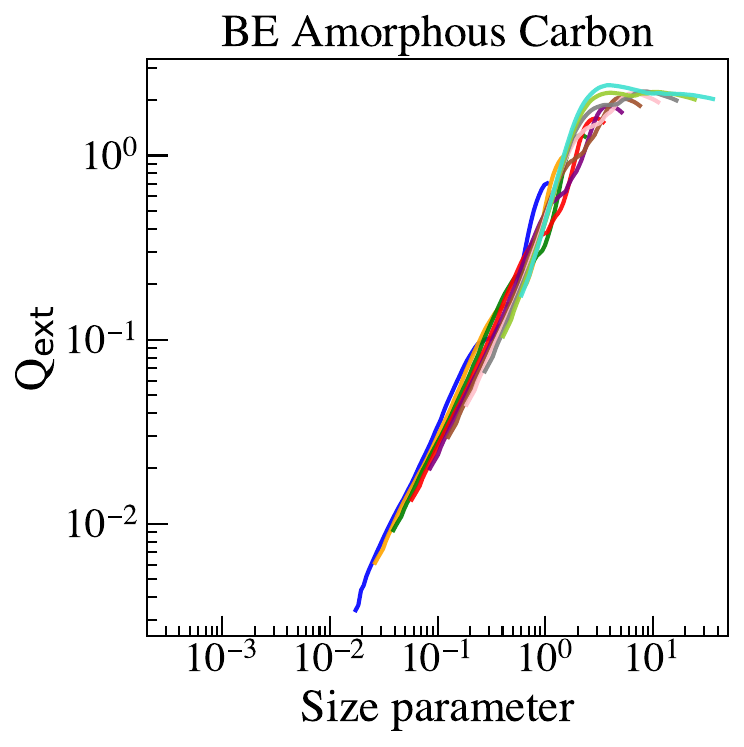}
\includegraphics[width=0.24\textwidth]{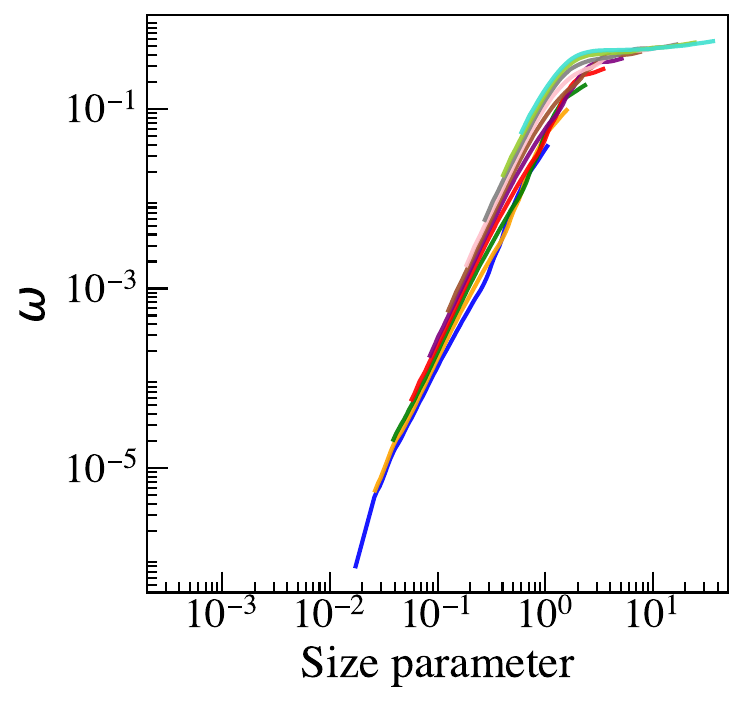}
\includegraphics[width=0.24\textwidth]{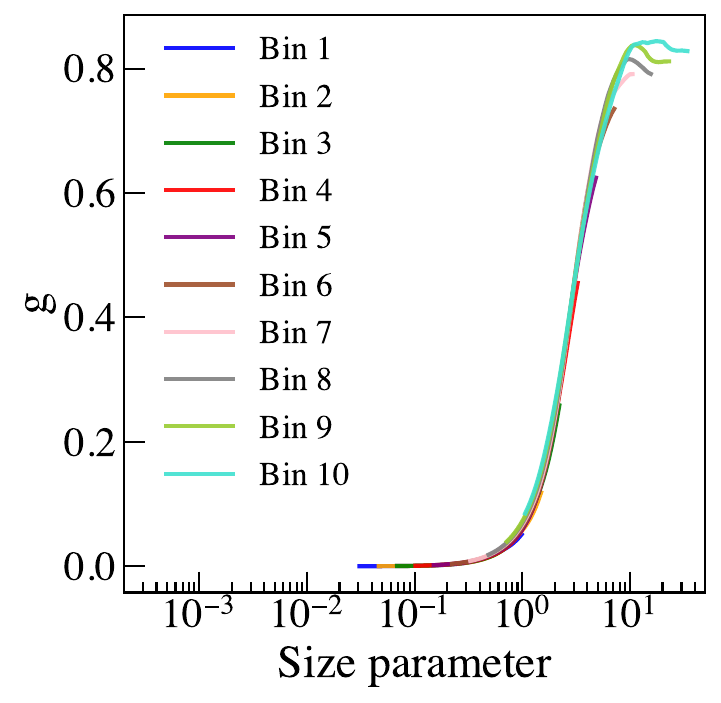}
\includegraphics[width=0.24\textwidth]{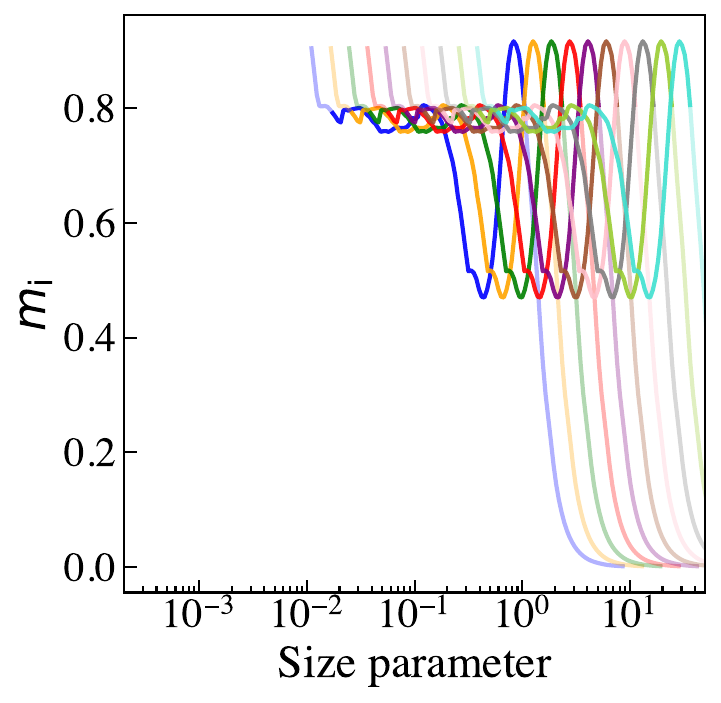}\\
\caption{The extinction efficiency (Q$_\mathrm{ext}$), single-scattering albedo ($\omega$), and asymmetry factor ($g$) obtained from the TAMUdust2020 database, together with the imaginary part of the refractive index for each dust species. The opaque region of the lines indicates the values used to calculate the opacities. The different colours refer to the bin size used for each calculation.}
\label{fig:TamuOut}
\end{figure*}

In Figure~\ref{fig:TamuOut}, we show the Q$_\mathrm{ext}$, $\omega$, and $g$ values obtained from the TAMUdust2020 database for each dust species. For size parameters $X\geq1$, the single-scattering albedo exhibits distinct oscillations as a function of $X$, originating from wave interference effects that occur when particle size becomes comparable to the wavelength. These ripple features are analogous to Mie resonances seen in spheres. 
The amplitude of these ripples is strongly influenced by the material's absorption: lower $m_\mathrm{i}$ allows interference effects to persist, while higher absorption dampens them. For $X\geq10$, minor oscillations in $g$ arise from residual interference effects and shape-dependent scattering effects of the hexahedral grains, despite orientation averaging.

As shown in Figure~\ref{fig:TamuOut}, the data required to estimate the opacity parameters are limited to a relatively narrow range of wavelengths (compared with the full wavelength range of \textsc{cloudy}). This limitation arises primarily from the refractive index available for each dust species and the space of parameters in the TAMUdust2020 database. 

To extend the opacity parameters across the full wavelength range used in \textsc{cloudy}, some assumptions were made. As previously mentioned, one assumption was that the opacity values in the large-particle regime were adopted from the spherical grains. On the other hand, for the small-particle regime, we adopted the power-law dependence observed in spherical grains following the behaviour of the intermediate-regime. However, the dust continuum emission is particularly sensitive to opacity in this regime, which corresponds to the MIR and FIR wavelengths. It is worth noting that linear interpolation was employed to ensure smooth transitions at the boundaries between the small-, intermediate-, and large- particle regimes.

In some cases, our input parameters ($m_\mathrm{r}$,$m_\mathrm{i}$, $\Psi$, $\lambda$,$D$) fell outside the parameter space covered by the TAMUdust2020 database. In such instances, scattering properties were calculated via extrapolation. These extrapolated values exhibited fluctuations to unrealistically small levels, manifesting as one or more sharp drops in the $SSA$ curves for $0.005 < X < 0.03$, and in the $g$ curves for $X< 0.03$. For $Q_\mathrm{ext}$, a small deviation is slightly visible at the smallest values of $X$ for grains sizes in bins 6, 7, and 8, which corresponds to the range $0.005 < X < 0.03$. These effects were accounted for during interpolation between the small- and intermediate-particle regimes.

For $\omega$, values within the affected range ($0.005 < X < 0.03$) were excluded and replaced by linear interpolation. For $g$, due to their proximity to zero these points were omitted. These considerations are observed in Figure~\ref{fig:TamuOut}.

\begin{table*}
\begin{center}
\begin{tabular}{lcccccccccc}
\hline                                     
Instrument          & \multicolumn{3}{|c|}{\textit{JWST}-MIRI}    & \textit{Spitzer}-MIPS  & \multicolumn{3}{|c|}{\textit{Herschel}-PACS} & \multicolumn{3}{|c|}{\textit{Herschel}-SPIRE}  \\ 
Filter             & F770W & F1130W & F1800W & MIPS\,24 & Blue & Green & Red & PSW & PMW & PLW\\ 
$\lambda_\mathrm{c}$ [$\mu$m] & 7.7  & 11.3         & 18.0         & 24.0     & 70.0     & 100.0     & 160.0     & 250.0      & 350.0      & 500.0\\ 
\hline
\multicolumn{8}{|l|}{Silicates}\\
\hline
B$_1$         & 1.0  & 1.0   & 1.0  & 1.0  & 1.0   & 1.1   & 1.1   & 1.1   & 1.1   & 1.1 \\
B$_2$         & 1.0  & 1.1   & 1.1  & 1.1  & 1.1   & 1.1   & 1.1   & 1.1   & 1.1   & 1.1 \\
B$_3$         & 1.0  & 1.1   & 1.2  & 1.1  & 1.1   & 1.1   & 1.1   & 1.1   & 1.1   & 1.1 \\
B$_4$         & 1.0  & 1.1   & 1.4  & 1.3  & 1.1   & 1.1   & 1.1   & 1.2   & 1.2   & 1.2 \\
B$_5$         & 1.0  & 1.0   & 1.0  & 1.1  & 1.3   & 1.4   & 1.4   & 1.5   & 1.5   & 1.5 \\
B$_6$         & 1.0  & 1.0   & 0.8  & 1.0  & 1.4   & 1.5   & 1.6   & 1.7   & 1.7   & 1.7 \\
B$_7$         & 1.0  & 1.0   & 0.9  & 1.1  & 1.5   & 1.6   & 1.7   & 1.7   & 1.7   & 1.7 \\
B$_8$         & 1.0  & 1.0   & 1.0  & 1.2  & 1.5   & 1.6   & 1.6   & 1.7   & 1.7   & 1.7 \\
B$_9$         & 1.0  & 1.0   & 1.0  & 1.0  & 1.5   & 1.7   & 1.8   & 1.8   & 1.9   & 1.9 \\
B$_{10}$      & 1.0  & 1.0   & 1.0  & 0.9  & 1.6   & 1.7   & 1.9   & 1.9   & 2.0   & 2.0 \\
MRN           & 1.0  & 1.0   & 1.1  & 1.1  & 1.3   & 1.4   & 1.5   & 1.6   & 1.6   & 1.6 \\
\hline
\multicolumn{8}{|l|}{Graphites}\\
\hline
B$_1$         & 1.0   & 0.5   & 0.3  & 0.5  & 1.2   & 1.4   & 1.6   & 1.9   & 2.0   & 2.1 \\
B$_2$         & 1.0   & 0.8   & 0.2  & 0.4  & 1.3   & 1.6   & 1.8   & 2.1   & 2.2   & 2.3 \\
B$_3$         & 1.0   & 0.9   & 0.2  & 0.4  & 1.4   & 1.7   & 1.8   & 2.2   & 2.3   & 2.3 \\
B$_4$         & 1.0   & 1.0   & 0.3  & 0.4  & 1.4   & 1.7   & 1.9   & 2.2   & 2.3   & 2.4 \\
B$_5$         & 1.0   & 1.0   & 0.4  & 0.4  & 1.5   & 1.7   & 1.9   & 2.2   & 2.3   & 2.4 \\
B$_6$         & 1.0   & 1.0   & 0.5  & 0.5  & 1.5   & 1.8   & 2.0   & 2.3   & 2.4   & 2.4 \\
B$_7$         & 1.0   & 1.0   & 0.7  & 0.4  & 1.6   & 2.0   & 2.2   & 2.6   & 2.8   & 2.9 \\
B$_8$         & 1.0   & 1.0   & 0.9  & 0.2  & 1.7   & 2.4   & 2.9   & 3.6   & 3.8   & 4.0 \\
B$_9$         & 1.0   & 1.0   & 1.0  & 0.3  & 1.7   & 2.3   & 2.8   & 3.5   & 3.7   & 3.9 \\
B$_{10}$      & 1.0   & 1.0   & 0.9  & 0.5  & 1.6   & 2.3   & 3.0   & 3.7   & 4.0   & 4.1 \\
MRN           & 1.0   & 0.9   & 0.3  & 0.4  & 1.5   & 1.9   & 2.3   & 2.8   & 3.0   & 3.1 \\
\hline
\multicolumn{8}{|l|}{AC Amorphous grains}\\
\hline
B$_1$         & 0.9   & 0.6   & 0.8   & 0.9  & 1.2   & 1.2   & 1.2   & 1.3   & 1.3   & 1.3 \\
B$_2$         & 1.0   & 0.5   & 0.7   & 0.8  & 1.2   & 1.2   & 1.3   & 1.3   & 1.4   & 1.4 \\
B$_3$         & 1.0   & 0.7   & 0.9   & 1.0  & 1.2   & 1.2   & 1.2   & 1.3   & 1.3   & 1.3 \\
B$_4$         & 1.0   & 0.8   & 0.9   & 1.0  & 1.3   & 1.3   & 1.3   & 1.4   & 1.4   & 1.4 \\
B$_5$         & 1.0   & 0.9   & 0.9   & 1.0  & 1.3   & 1.3   & 1.3   & 1.4   & 1.4   & 1.4 \\
B$_6$         & 1.0   & 1.0   & 1.0   & 1.1  & 1.4   & 1.3   & 1.4   & 1.4   & 1.4   & 1.4 \\
B$_7$         & 1.0   & 1.0   & 1.1   & 1.2  & 1.4   & 1.4   & 1.4   & 1.4   & 1.4   & 1.4 \\
B$_8$         & 1.0   & 1.0   & 1.2   & 1.4  & 1.5   & 1.4   & 1.4   & 1.4   & 1.4   & 1.4 \\
B$_9$         & 1.0   & 1.0   & 1.1   & 1.4  & 1.5   & 1.4   & 1.5   & 1.5   & 1.5   & 1.5 \\
B$_{10}$      & 1.0   & 1.0   & 1.0   & 1.5  & 1.6   & 1.5   & 1.5   & 1.5   & 1.5   & 1.5 \\
MRN           & 1.0   & 0.7   & 0.8   & 1.0  & 1.4   & 1.3   & 1.4   & 1.4   & 1.4   & 1.4 \\
\hline
\multicolumn{8}{|l|}{BE Amorphous grains}\\
\hline
B$_1$         & 1.0   & 0.6   & 0.7   & 0.8  & 1.2   & 1.3   & 1.4   & 1.5   & 1.5   & 1.5 \\
B$_2$         & 1.0   & 0.6   & 0.6   & 0.8  & 1.2   & 1.4   & 1.4   & 1.5   & 1.5   & 1.6 \\
B$_3$         & 1.0   & 0.7   & 0.6   & 0.8  & 1.3   & 1.4   & 1.5   & 1.5   & 1.6   & 1.6 \\
B$_4$         & 1.0   & 0.9   & 0.6   & 0.8  & 1.3   & 1.4   & 1.5   & 1.6   & 1.6   & 1.6 \\
B$_5$         & 1.0   & 0.9   & 0.7   & 0.8  & 1.3   & 1.4   & 1.5   & 1.6   & 1.6   & 1.6 \\
B$_6$         & 1.0   & 1.0   & 0.8   & 0.9  & 1.4   & 1.5   & 1.6   & 1.6   & 1.6   & 1.6 \\
B$_7$         & 1.0   & 1.0   & 0.8   & 1.0  & 1.4   & 1.5   & 1.6   & 1.6   & 1.7   & 1.7 \\
B$_8$         & 1.0   & 1.0   & 0.9   & 1.1  & 1.5   & 1.5   & 1.6   & 1.7   & 1.7   & 1.7 \\
B$_9$         & 1.0   & 1.0   & 1.0   & 1.1  & 1.5   & 1.6   & 1.6   & 1.7   & 1.7   & 1.7 \\
B$_{10}$      & 1.0   & 1.0   & 1.0   & 1.2  & 1.5   & 1.6   & 1.7   & 1.7   & 1.7   & 1.7 \\
MRN           & 1.0   & 0.8   & 0.7   & 0.8  & 1.4   & 1.5   & 1.6   & 1.6   & 1.6   & 1.7\\
\hline
\end{tabular}
\caption{Ratio of synthetic fluxes derived from hexahedral and spherical grain models, F$_\mathrm{hex}$/F$_\mathrm{sph}$, across a range of infrared observational bands: \textit{JWST}-MIRI, \textit{Spitzer}-MIPS, and \textit{Herschel}-PACS and -SPIRE. Each row corresponds to a different dust size distribution bin (B$_1$--B$_{10}$) or the standard MRN distribution, and values are listed separately for silicates, graphite, and amorphous carbon (AC and BE) compositions.}
\label{tab:Obsbands}
\end{center}
\end{table*}

\subsection{Modelling nebulae around evolved stars}

SED analysis of PNe and WRN using photoionization modelling enables us to reconstruct the final mass-loss episodes and dust production during the previous AGB, RSG, YSG or LBV phase of stellar evolution. For example, \citet{2025Astro...4....2T} have traced the evolution of the SED through the AGB, post-AGB and PN stages of 137 sources in the Galaxy and the Magellanic Clouds. In the AGB stage, the dust and stellar contributions overlap because of physical proximity, high dust temperature and low stellar photospheric temperature. As the central star evolves and heats up, the dust shell detaches and cools, and the post-AGB star and shell contributions to the SED can be distinguished. In the PN stage, there are three components to the SED: the stellar contribution in the optical and UV, the nebular gas contribution in the optical and near IR, and the expanding dust shell at mid and far IR wavelengths. A similar evolutionary history applies to WRN, where a high-mass loss RSG, YSG or LBV stage precedes the hot WR star. Photoionization modelling of PN and WRN emission-line spectra and SEDs enables the self-consistent determination of the stellar ionizing luminosity and effective temperature, together with the gas density and abundances in the nebula, the dust species, dust-to-gas mass ratio and the grain size distribution. This large parameter space means that such photoionization models are highly degenerate and any additional constraints are useful in restricting the ranges and number of free parameters.

Tight constraints on the production of silicate grains during the AGB phase of intermediate mass stars (4--7~$M_\odot$) have been obtained by comparing the results from stellar evolution and dust formation modelling with the observed SEDs of a sample of solar metallicity, intermediate-mass, single AGB stars \citep{2023A&A...670A..97M}. Dust yields in the range 0.012--0.025~$M_\odot$ were found for progenitors in the range 4--7~$M_\odot$ by \citet{2023A&A...670A..97M} but substantially lower yields (0.002--0.006~$M_\odot$) were predicted by \citet{2018MNRAS.475.2282V} from theoretical models. For lower-mass (1.5--3~$M_\odot$) carbon-rich stars, both solid carbon and SiC grains are produced, with predicted yields in the ranges 0.001--0.02~$M_\odot$ and $3\times10^{-4}$ to $3\times10^{-3} M_\odot$, respectively \citep{2018MNRAS.475.2282V}. A comparison of synthetic SEDs with \textit{Spitzer} IR spectra for a sample of carbon-rich AGB stars in the LMC shows that 80\% of the dust formed is solid carbon, with 10--20\% of this attributed to graphite and the rest to amorphous carbon. The remaining 20\% of the total dust mass is due to SiC \citep{2021A&A...647A..69M}. 

Scattered-light imaging of the dust shells surrounding three intermediate-mass AGB stars has revealed silicate grain sizes up to 0.3~$\mu$m in radius \citep{2012Natur.484..220N}. Radiative transfer modelling of the 70~$\mu$m and 160~$\mu$m emission 
from the AGB wind--ISM interaction regions of a sample of AGB and RSG stars shows that grain sizes for carbon-rich sources are preferentially large ($\sim 2\mu$m), whereas the derivation of dust properties for the oxygen-rich sources is not straightforward \citep{2022A&A...663A..64M}. However, once the central star enters the PN phase, the hardening of the stellar radiation field and sputtering of dust grains by the hot phase of the nebular gas will modify the size distribution of the grains and the dust-to-gas mass ratio. Constraining the PN dust properties using photoionization modelling can help to determine the timescales over which the dust destruction processes have acted. However, as we found in Section~\ref{sec:results}, the inferred dust mass and size distribution are highly dependent on the dust optical properties and the shape assumed for the grains.
  
Modelling the SED in the far-IR is particularly challenging, as this emission originates from the coldest dust grains in the nebula. To reproduce this component, under the assumption that grains lie at similar distances from the central star, it is often necessary to include large grains (with sizes $\gtrsim$ 1~$\mu$m).
 
Alternatively, some studies have invoked dust-to-gas mass ratios higher than the canonical value of 1/100 to match the observed far-IR fluxes. However, both options could conflict with physical constraints from dust formation and mass-loss history. For example, \citet{2016A&A...588A..92V} reproduced the far-IR SED of the WRN M\,1-67 with a pure dust model in which 95\% of the dust mass corresponds to grains with sizes in the range 2--10~$\mu$m, and suggested that the progenitor must have been a RSG. However, \citet{2020MNRAS.497.4128J} were able to reproduce the far-IR SED of the same nebula with a multi-shell photoionization model including gas and dust and required a maximum grain size of only 0.9~$\mu$m. They proposed a richer evolutionary history involving a common envelope binary scenario. These contrasting results highlight the degeneracy inherent in far-IR SED modelling, as different combinations of grain sizes, dust masses and nebular properties can reproduce similar SEDs. In particular, models with micron-sized grains may imply dust masses that are difficult to reconcile with plausible mass-loss histories. Because of this degeneracy, the SED alone cannot uniquely discriminate between different scenarios, and additional observational constraints, such as polarization measurements, are required, motivating the exploration of alternative explanations.

A third approach involves varying the grain geometry. Non-spherical shapes such as ellipsoidal, fluffy, or irregular shapes, can enhance far-IR emission without requiring large grain sizes or high dust abundances. In this context, hexahedral grains provide a more realistic approximation of irregular dust morphologies and offer a physically motivated means of reconciling models with observed far-IR photometry.

Dust opacities in two different spectral bands are important for determining the infrared spectrum:
\begin{enumerate}
\item
Ultraviolet opacities determine the absorption optical depth of the dust shell
and hence the total luminosity of the infrared reprocessed emission.
\item
Infrared opacities determine the cooling efficiency of the grains,
and hence their radiative equilibrium temperature,
which in turn determines the spectral shape of the infrared radiation.
\end{enumerate}

As discussed in Section~\ref{sec:results}, implementing hexahedral dust grains results in a shift of 
the location of the IR peak in the spectral energy distribution. This peak is particularly significant, as it serves as a diagnostic of the maximum grain size contributing to the IR dust emission. The importance of this grain size lies in its connection to the physical processes of dust formation \citep[e.g.][]{Kochanek2011}, which are governed by the evolution of the star-nebula system.

\subsubsection{Synthetic fluxes and temperature profiles}

Table~\ref{tab:Obsbands} presents the ratios of synthetic fluxes from our \textsc{cloudy} models comparing hexahedral and spherical grains across ten IR bands, spanning wavelengths from 8 -- 500~$\mu$m, where the dust continuum is prominent. The largest discrepances between models are observed at longer wavelengths. For example, in the 500~$\mu$m band, the flux for graphite in the irregular-grain model is up to four times higher than in the spherical-grain model. Similarly, for silicates, the flux can be up to twice as high -- differences that are clearly non-negligible. For amorphous carbon grains, the enhancement is present with flux ratios between 1.5 and 1.7 compared to the spherical-grain model. 

For the case of silicates, non-spherical grains show enhanced emission 
(flux ratios greater than unity)
for all wavelengths longer than 10~$\mu$m (see Fig.~\ref{fig:ModelSil}). 
Graphite grains behave differently,
showing flux ratios less than unity at shorter mid-infrared wavelengths ($< 70~\mu$m)
but greater than unity in the far infrared,
while amorphous carbon shows a mixture of these two behaviors.

This can be understood with reference to the changes in the dust opacity over the ultraviolet and infrared wavebands, as mentioned above.
For silicates, the most important factor is the increased opacity at UV wavelengths, 
especially for the larger grains (Fig.~\ref{fig:opacsil}), 
which leads to a general increase in the luminosity of the dust shell.
For graphite, the most important factor is the 
increased opacity at IR wavelengths (Fig.~\ref{fig:opacgrap}),
which leads to lower grain temperatures
and a shift of the dust shell emission towards longer wavelengths.
 
\begin{figure*}
\centering
\includegraphics[width=0.8\textwidth]{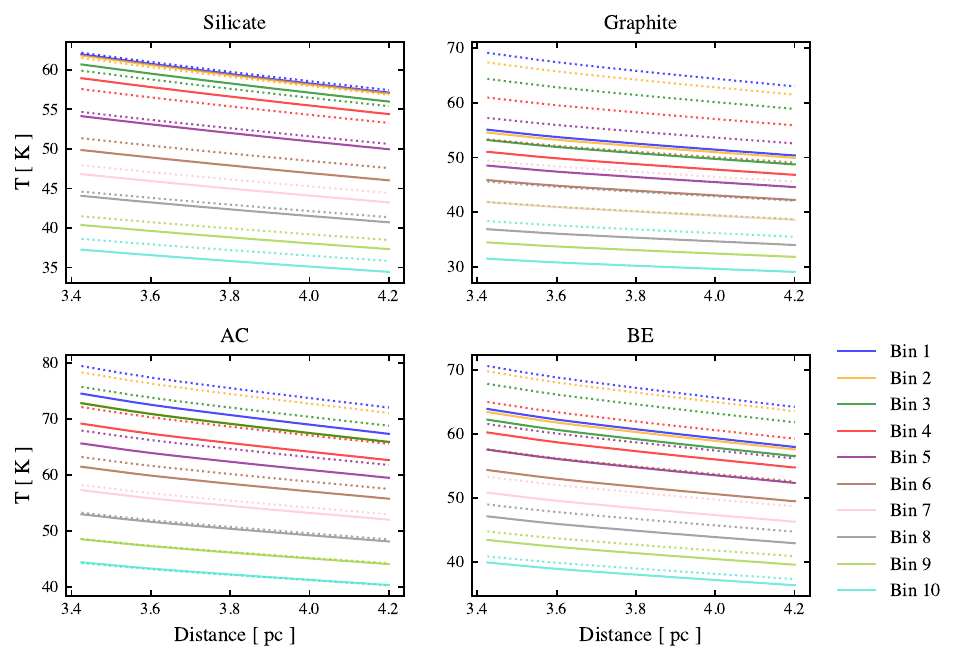}
\caption{Dust grain temperature profiles from our photoionization models, computed using the MRN size distribution divided into 10 size bins for silicate, graphite and amorphous carbon grains. Solid lines represent models with hexahedral grains, while dotted lines correspond to spherical grain models.}
\label{fig:dusttemp}
\end{figure*}

Figure~\ref{fig:dusttemp} shows the dust temperature as a function of the radial distance from the star for the models employing the MRN size distribution (0.005 -- 0.25~$\mu$m) for each dust species. The figure compares results for spherical and hexahedral grain geometries, illustrating that hexahedral grains generally reach lower temperatures than their spherical counterparts, as expected due to the differences in their opacities.

As shown in Figure~\ref{fig:dusttemp}, the most significant difference occurs for graphite. In the innermost regions of the nebula, the smallest spherical grains reach temperatures of approximately 70~K, whereas the corresponding hexahedral grains attain temperatures roughly 15~K lower. For amorphous carbon grains, the temperature difference is more noticeable among the smallest grains, but the temperatures converge for larger grain sizes. In the case of silicate grains, although the temperature difference increases with grain size, the overall disparity remains relatively modest.

\subsection{Impact on polarization analysis}

Considering the geometry (as well as the porosity) of dust grains is particularly relevant in the context of polarimetric analysis across near-infrared (NIR) and FIR wavelengths. Indeed, in the NIR, the linear polarization is dominated by scattering (by small grains), and the polarization patterns are therefore revealing the distribution of dusty environments. As dust grains with distinct shapes will produce distinct polarization angles and polarization degrees \citep{Olofsson2024}, an accurate analysis and description of shells or disks around astrophysical objects such as protostars or evolved stars, for example, requires an adequate description of the dust characteristics. \\
In the FIR down to the millimeter (mm) range, the origin of the polarization is primarily due to emission, but also scattering, from non-spherical aligned dust grains. The processes responsible for the dust alignment can be diverse, although radiative torques are generally invoked \citep{Tram2022}. While the alignment efficiency of the grains is sensitive to their composition, we note that it is also partly related to their shape. This has been discussed, for example, by \citet{Lazarian2007}, \citet{Hoang2008} and \citet{Herranen2019} in the case of Radiative Torque Alignment (RAT). Therefore, since the magnetic field distribution in various astrophysical objects is obtained from the polarized dust emission \citep{Planck2020,Sabin2020}, it is also dependent on the grain geometry.

To explore the impact of dust grain geometry on polarization, a broader study covering multiple wavelengths will be reported in future work. This will include comparisons between photoionization models of evolved stars using different dust grain shapes, including the dust description outlined in this article.\\

\section{Summary}
\label{sec:summary}
We have used the TAMUdust2020 database to obtain the single-scattering properties of hexahedral silicate and carbonaceous grains with size parameters relevant to dust in photoionized nebulae around evolved stars. These size parameters correspond to grains with sizes in the range 0.005--0.25~$\mu$m and wavelengths in the NIR and MIR. Outside of this wavelength range, standard asymptotic power-law behaviour was assumed in the small- and large-particle limits.

The TAMUdust2020 values were used to calculate the absorption and scattering cross sections per H nucleon as a function of wavelength as input to the \textsc{cloudy} photoionization code. We constructed a two-shell photoionized model representing circumstellar material around a hot central star, mimicking mass-loss episodes typical of evolved stars. We then performed a set of numerical experiments to compare the effects of spherical and hexahedral grains on the emergent SED, exploring  silicate and carbonaceous grains and different grain sizes within the MRN dust distribution.

Our main findings are as follows:

\begin{enumerate}
    \item Models incorporating irregular hexahedral grains produce a shift of the dust continuum towards longer wavelengths and result in enhanced far-IR fluxes compared to spherical grain models.
    \item This effect can account for the observed far-IR emission without requiring large grain sizes or high dust-to-gas mass ratios.
    \item Photoionized models using spherical grains systematically underestimate the far-IR emission unless more dust mass is added, potentially biasing estimates of dust content and mass-loss history in evolved star nebulae.
    \item The effects on the dust continuum are more sensitive to the dust absorptivity, and thus depend strongly on the grain composition. This is particularly evident for graphite grains, which exhibit more pronounced changes in the continuum compared to the other dust species.
\end{enumerate}

We emphasize that the primary aim of this work is to demonstrate, in a general framework, the differences between arising from the inclusion of spherical versus more realistic grain shapes in photoionization modelling. In future work, this approach will be applied to a specific object, as modelling real systems requires a detailed treatment, particularly because the distribution of nebular material and the stellar spectrum must be tailored to each case. In this context, our results suggest that the adoption of more realistic, irregular grain shapes (such as hexahedral-shaped grains) can alleviate some of the modelling challenges in interpreting IR observations of evolved star environments. They also offer new pathways for constraining dust content and the evolutionary history of massive stars through IR diagnostics.

Future improvements will require more comprehensive measurements of dust optical properties across the full range of astrophysically relevant wavelengths. Additionally, a better understanding of the asymptotic behaviour of absorption and scattering efficiencies in the small- and large-particle regimes is needed. Finally, this framework may also be extended to investigate polarization arising from non-spherical grains.

\section*{Acknowledgements}

This work was supported by Universidad Nacional Aut\'onoma de M\'exico through the DGAPA-UNAM postdoctoral fellowships Program. PJH, SJA, and WJH acknowledge financial support from UNAM PAPIIT project IN109823. PJH, DG, OM and JM acknowledge support by grants PID2021-123370OB-100 AEI/10.13039/501100011033/FEDER,  PID2024-156713OB-I00/AEI/FEDER, and  Severo Ochoa CEX2021-001131-S funded by MCIN/AEI/ 10.13039/501100011033. LS and PJH acknowledge financial support from UNAM PAPIIT project IN107625. PJH thanks J.A. Toal\'{a} and M.A. Guerrero for helpful discussions. This research has made use of the Astrophysics Data System, funded by NASA under Cooperative Agreement 80NSSC21M00561.

\section*{Data Availability}

The data underlying this article will be shared on reasonable request to the corresponding author.

\section*{Software}
Software: Cloudy \citep{Ferland2017}, Astropy \citep{2013A&A...558A..33A,2018AJ....156..123A}, TAMUdust2020 \citep{Saito2021} Matplotlib \citep{2007CSE.....9...90H}, NumPy \citep{2011CSE....13b..22V}, SciPy \citep{2020NatMe..17..261V}.


\bibliographystyle{mnras}
\bibliography{biblio} 



\appendix

\section{Additional figures}

Here we present the comparison of the emergent spectra from \textsc{cloudy} models using spherical and hexahedral-shaped grains, for bin sizes 2 through 9 as listed in Table~\ref{tab:bines}. Results are shown for silicate, graphite, and amorphous carbon grain compositions.

\begin{figure*}
\centering
\includegraphics[width=0.24\textwidth]{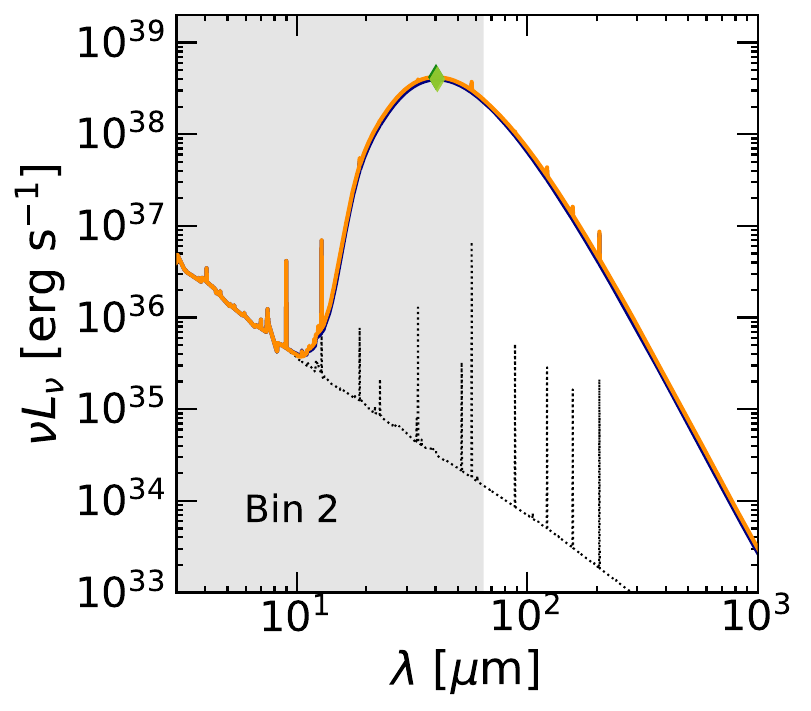}
\includegraphics[width=0.24\textwidth]{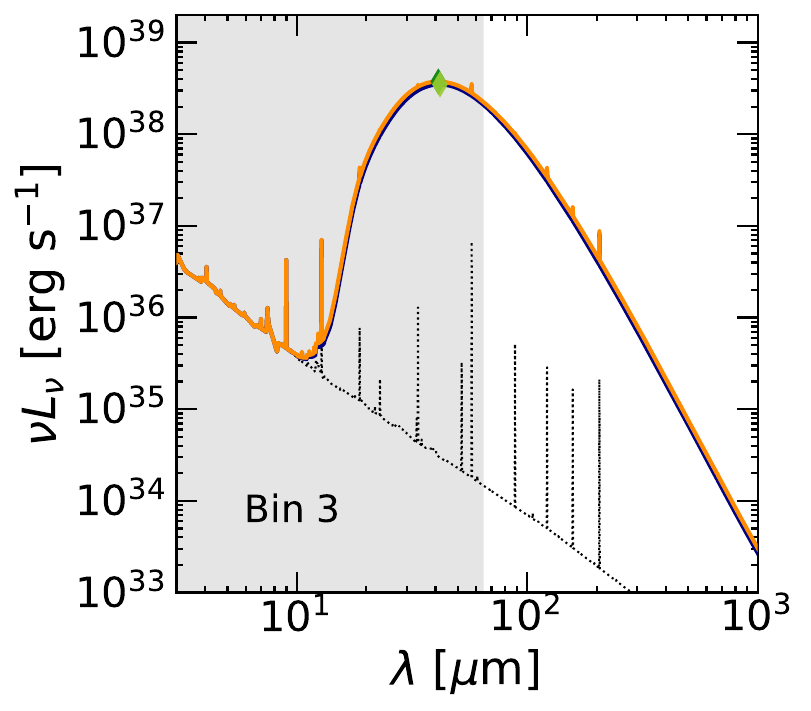}
\includegraphics[width=0.24\textwidth]{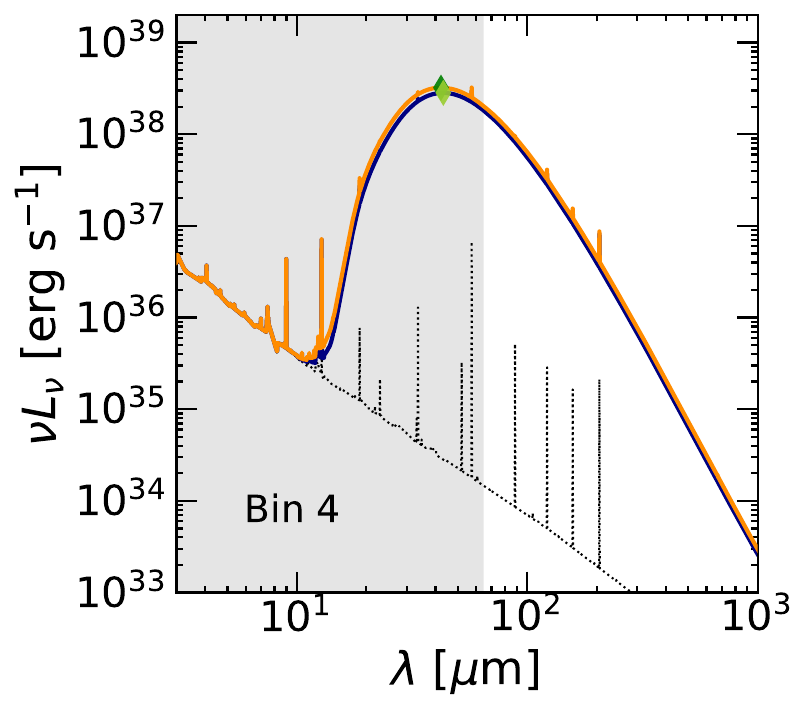}
\includegraphics[width=0.24\textwidth]{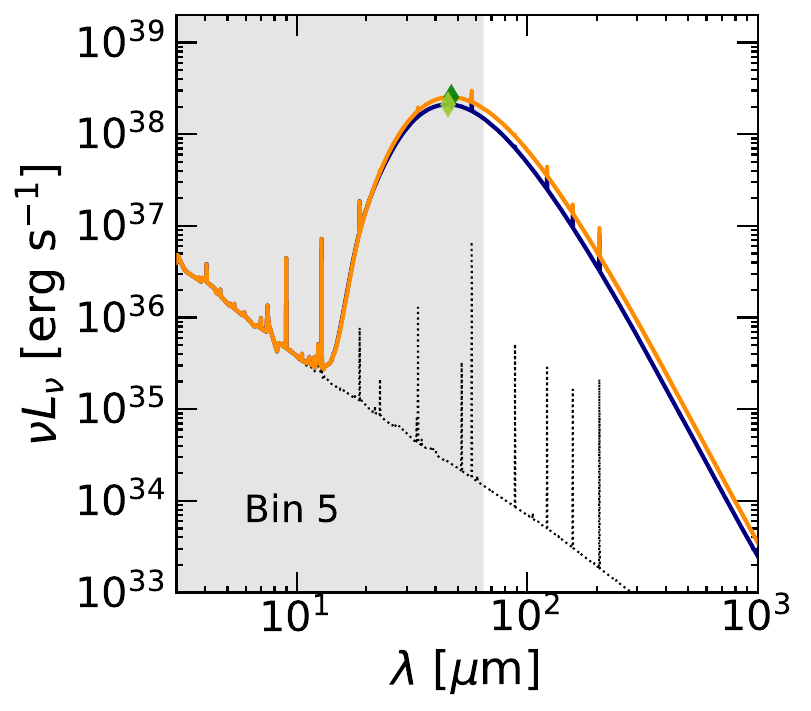}\\
\includegraphics[width=0.24\textwidth]{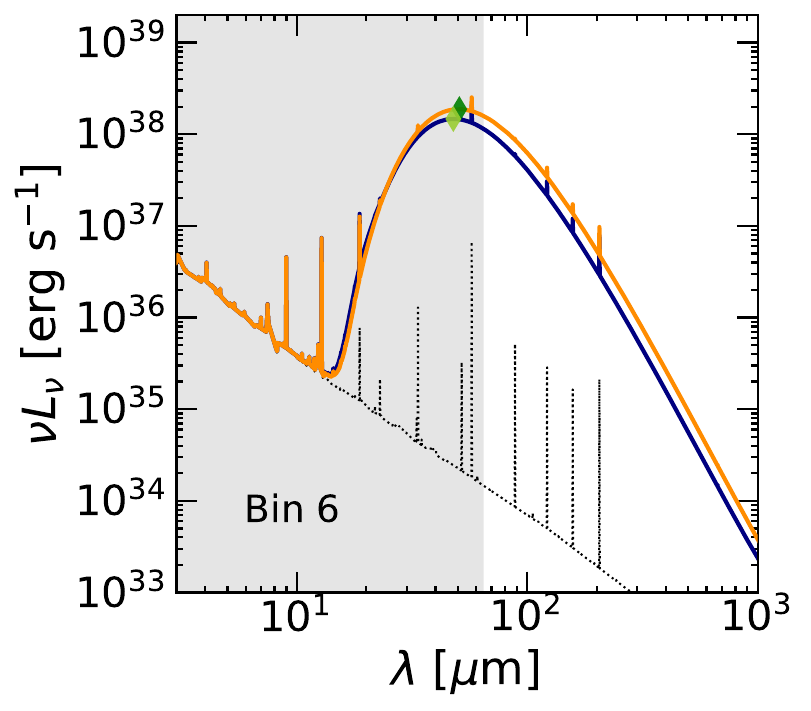}
\includegraphics[width=0.24\textwidth]{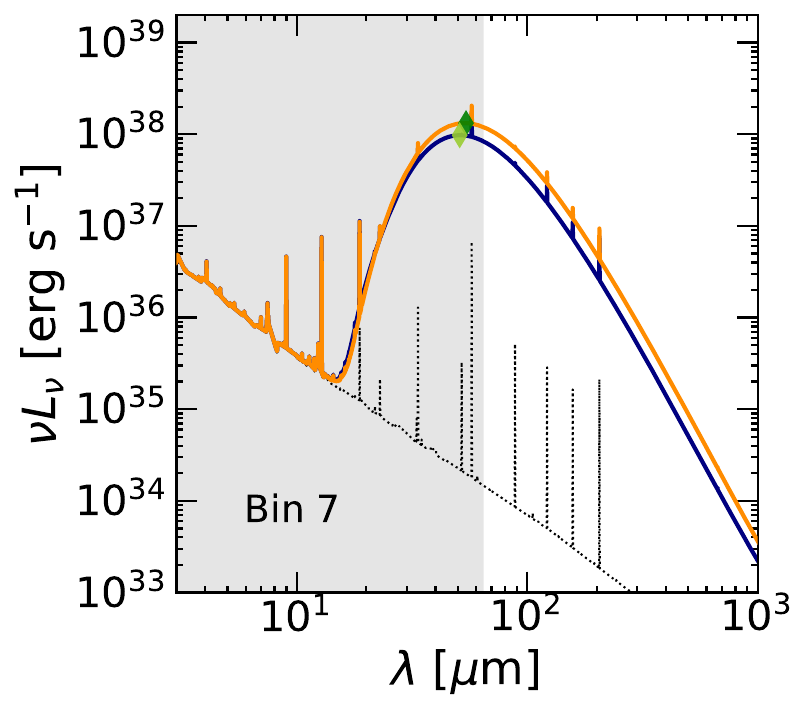}
\includegraphics[width=0.24\textwidth]{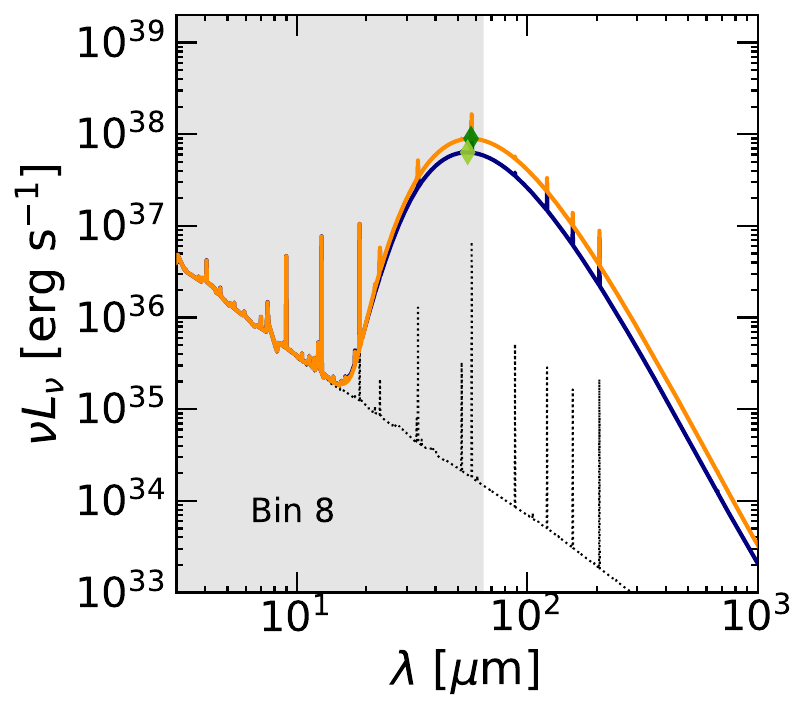}
\includegraphics[width=0.24\textwidth]{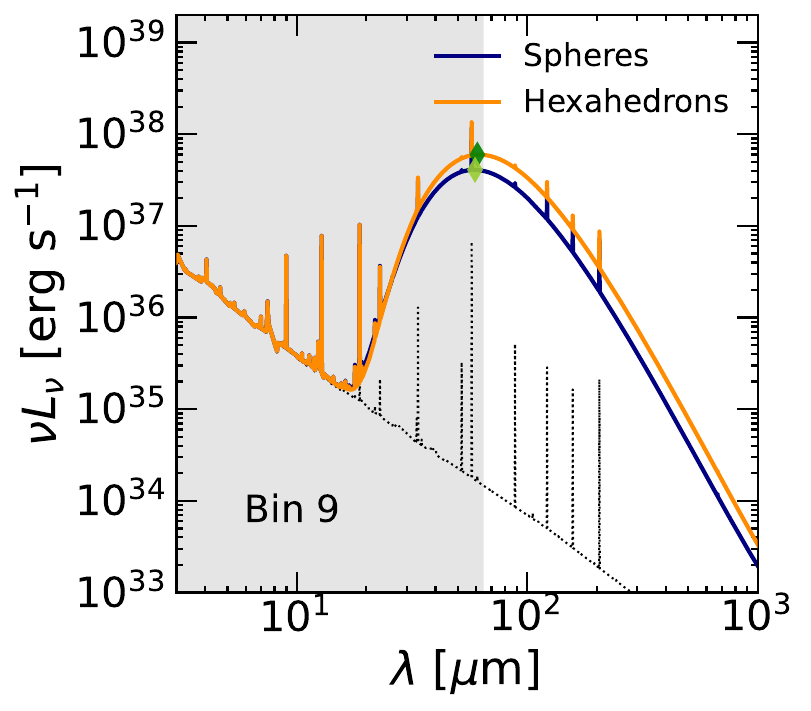}
\caption{Spectra from \textsc{cloudy} photoionization models with silicate grains, comparing spherical (blue line) and hexahedral-shaped (orange line) grains. Each panel shows models using only the grain sizes corresponding to bins 2 through 9 from Table~\ref{tab:bines}. The contribution of the stellar spectrum is shown in black, and green diamonds mark the IR continuum peak of each model. The shaded grey region indicates wavelengths where opacities were derived from TAMUdust2020 output data.}
\label{fig:ModelSilapp}
\end{figure*}

\begin{figure*}
\centering
\includegraphics[width=0.24\textwidth]{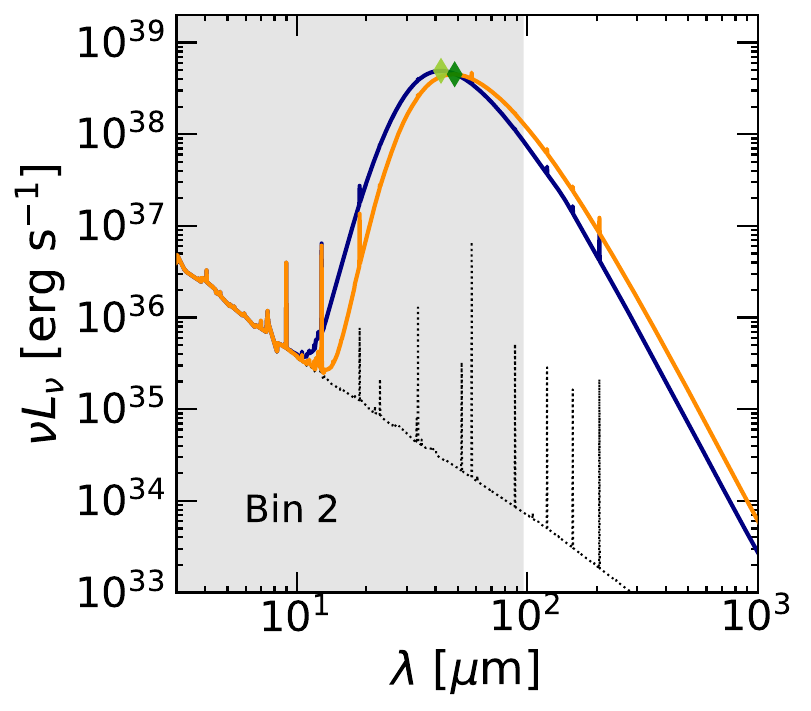}
\includegraphics[width=0.24\textwidth]{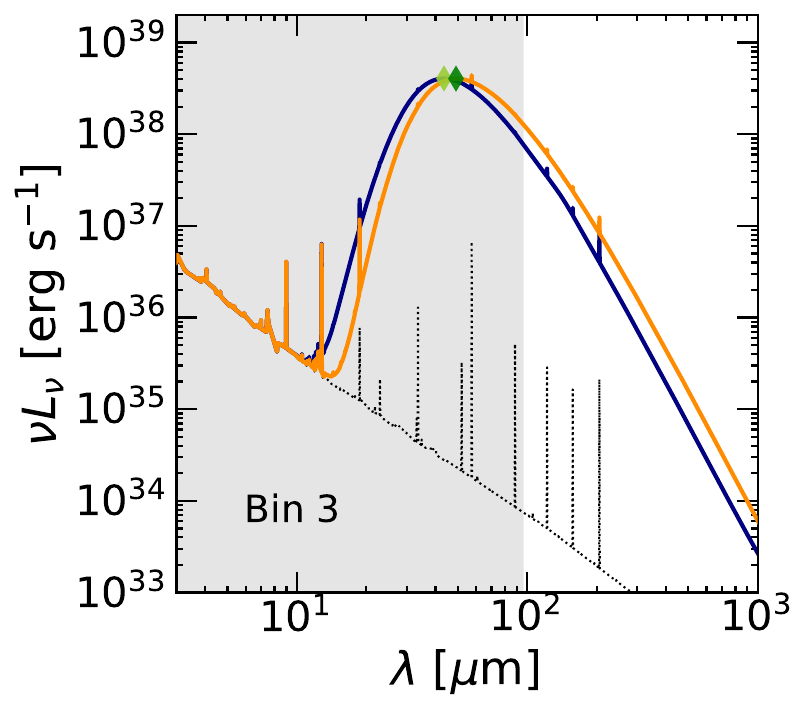}
\includegraphics[width=0.24\textwidth]{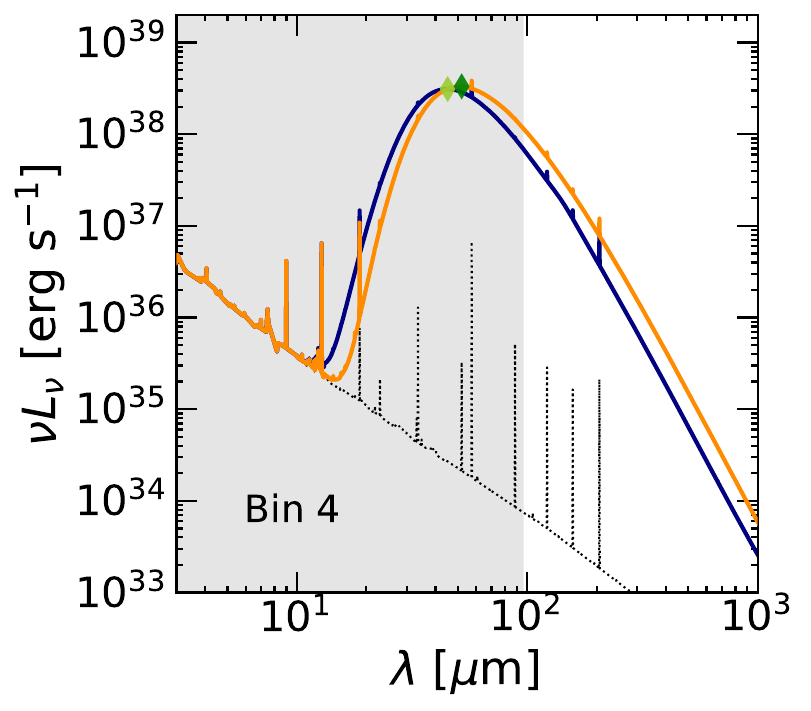}
\includegraphics[width=0.24\textwidth]{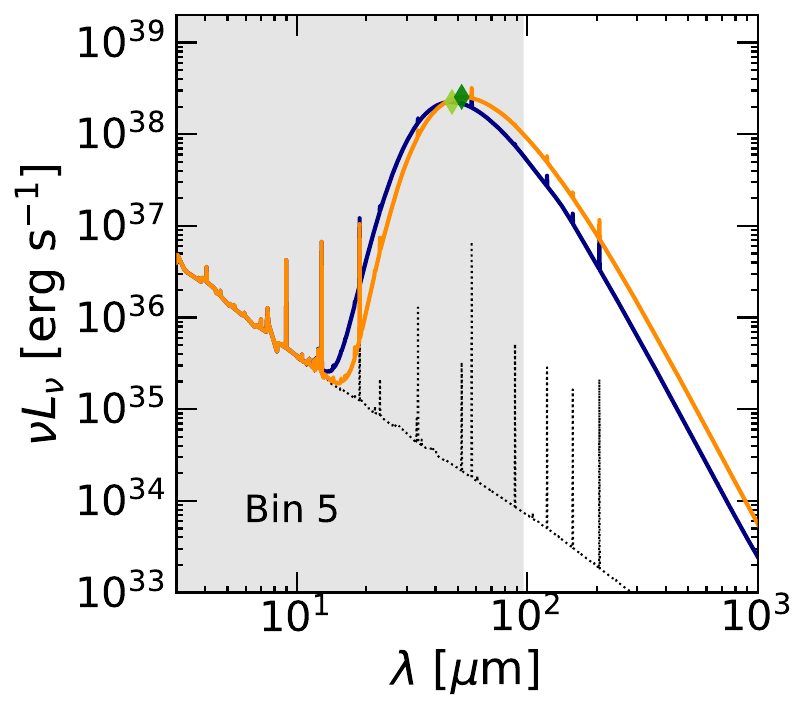}\\
\includegraphics[width=0.24\textwidth]{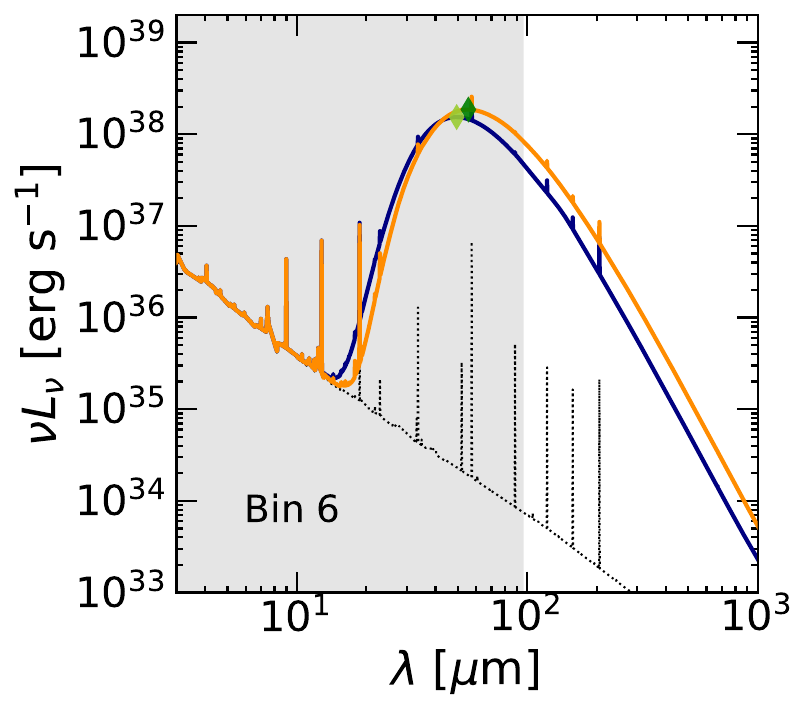}
\includegraphics[width=0.24\textwidth]{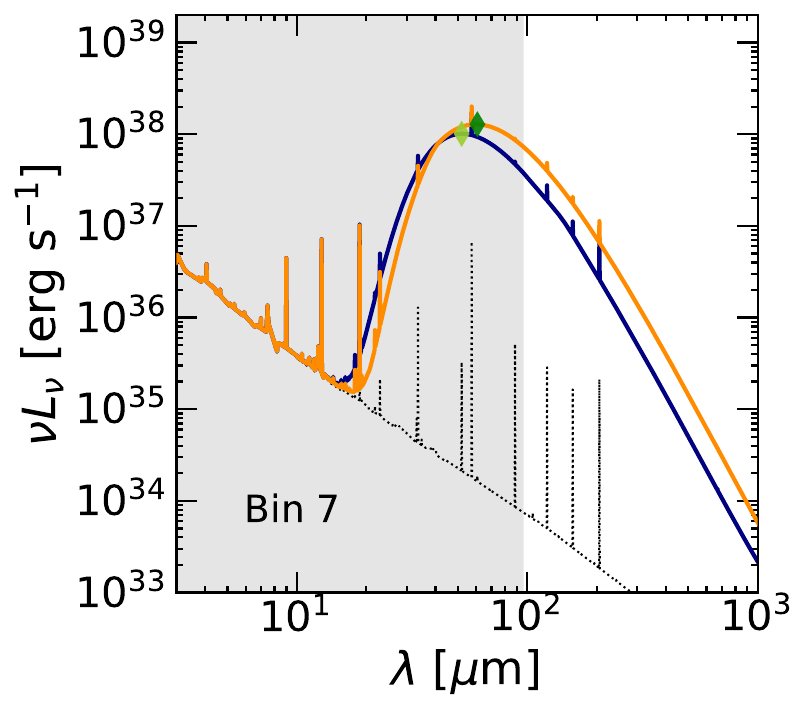}
\includegraphics[width=0.24\textwidth]{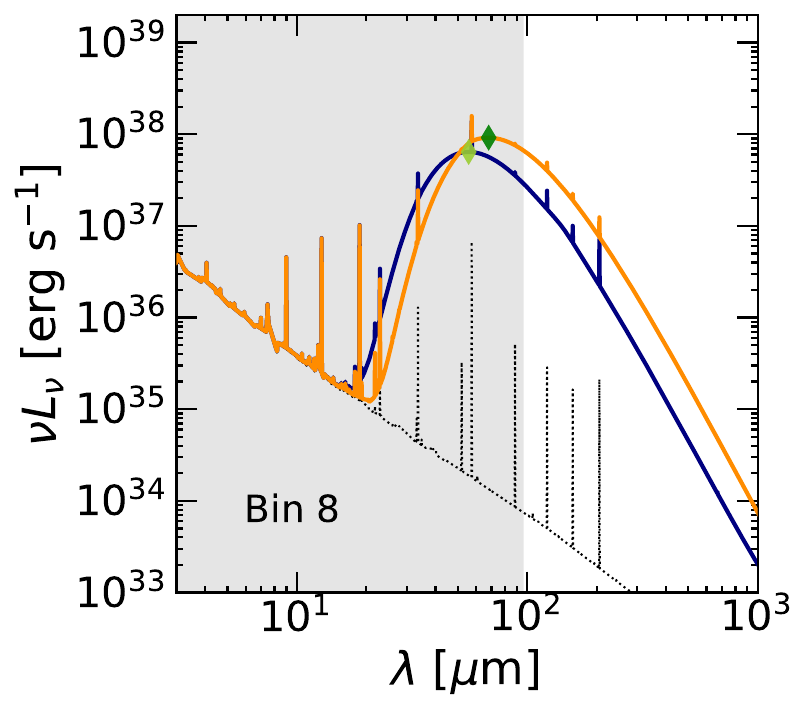}
\includegraphics[width=0.24\textwidth]{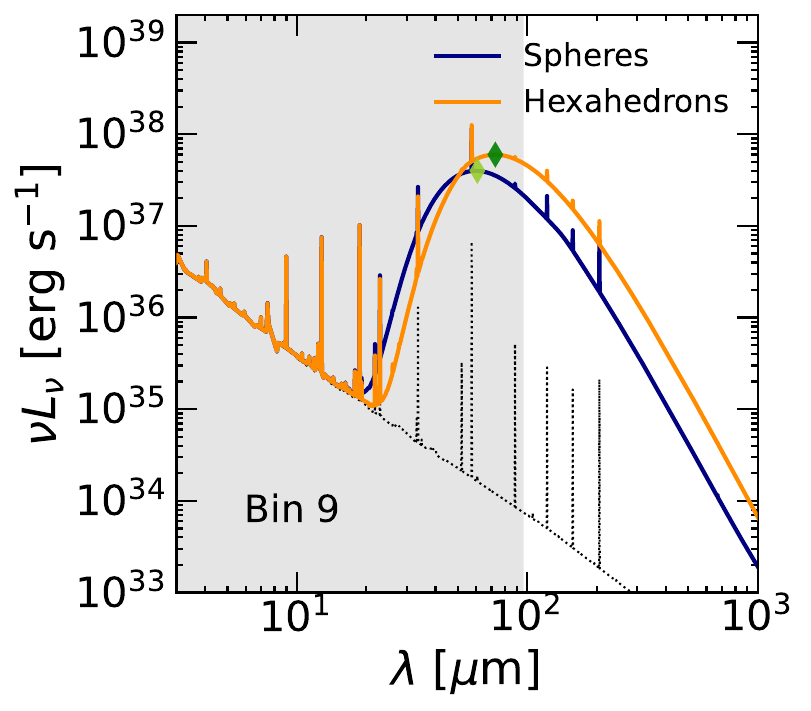}
\caption{Spectra from \textsc{cloudy} photoionization models with graphite grains, comparing spherical (blue line) and hexahedral-shaped (orange line) grains. Each panel shows models using only the grain sizes corresponding to bins 2 through 9 from Table~\ref{tab:bines}. The contribution of the stellar spectrum is shown in black, and green diamonds mark the IR continuum peak of each model. The shaded grey region indicates wavelengths where opacities were derived from TAMUdust2020 output data.}
\label{fig:ModelGrapapp}
\end{figure*}

\begin{figure*}
\centering
\includegraphics[width=0.24\textwidth]{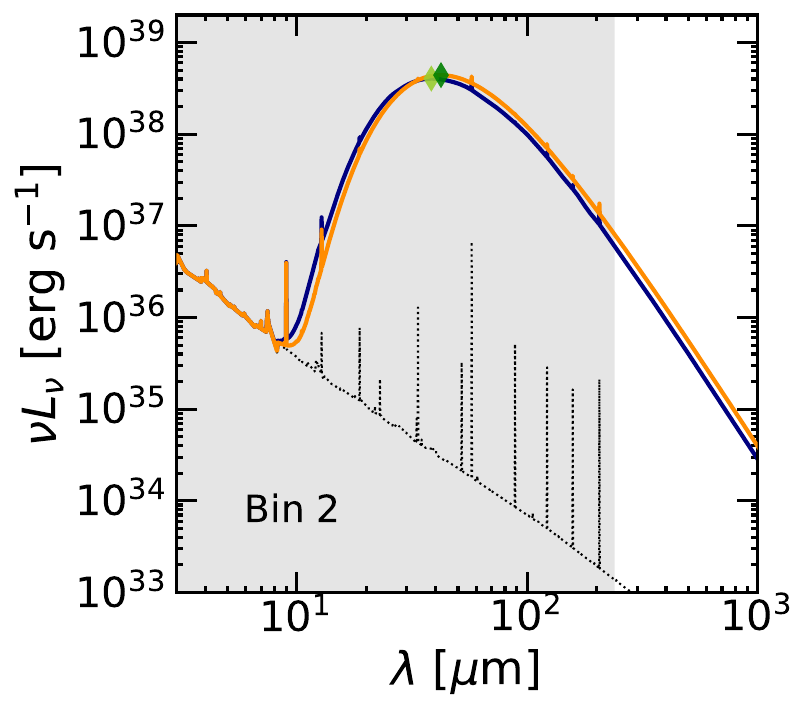}
\includegraphics[width=0.24\textwidth]{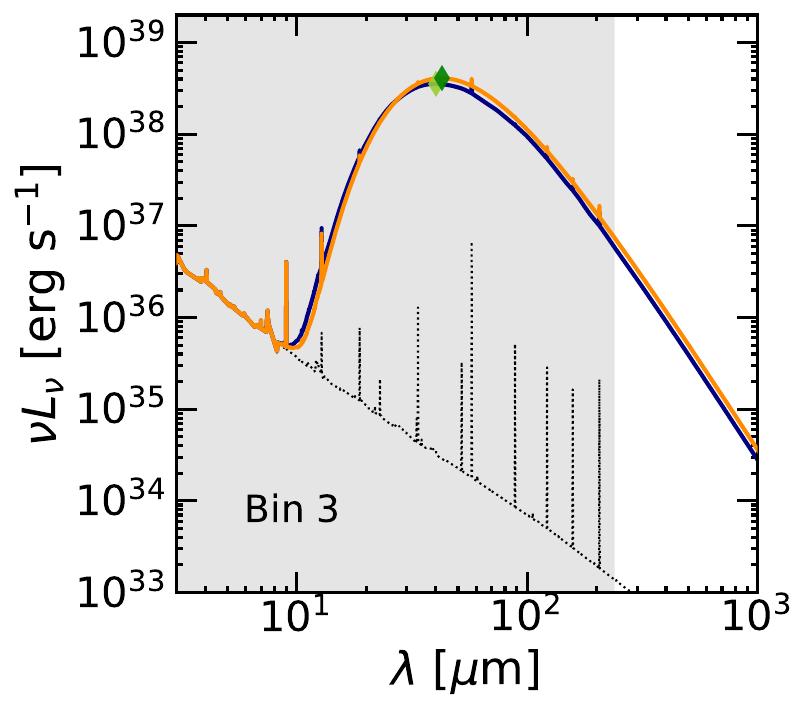}
\includegraphics[width=0.24\textwidth]{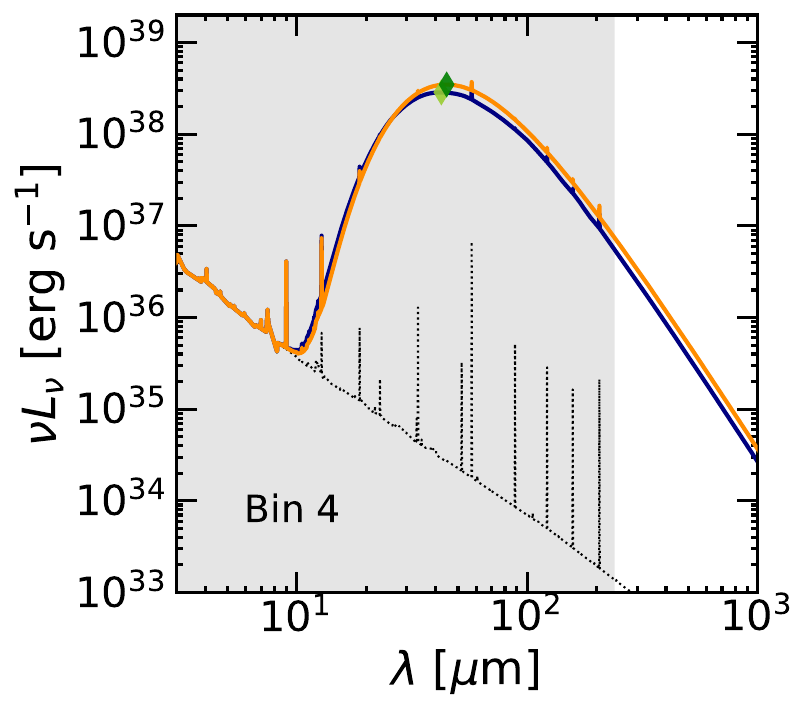}
\includegraphics[width=0.24\textwidth]{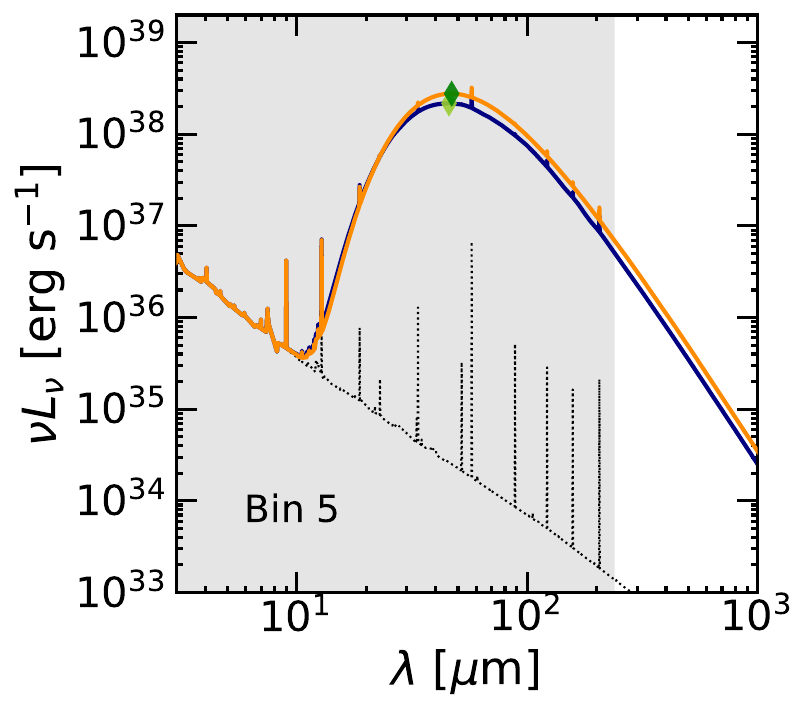}\\
\includegraphics[width=0.24\textwidth]{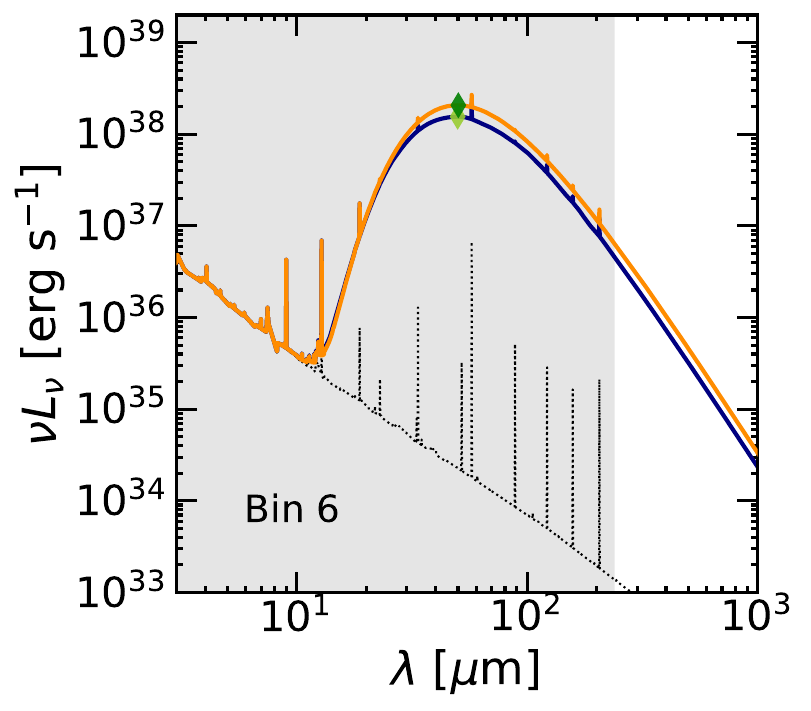}
\includegraphics[width=0.24\textwidth]{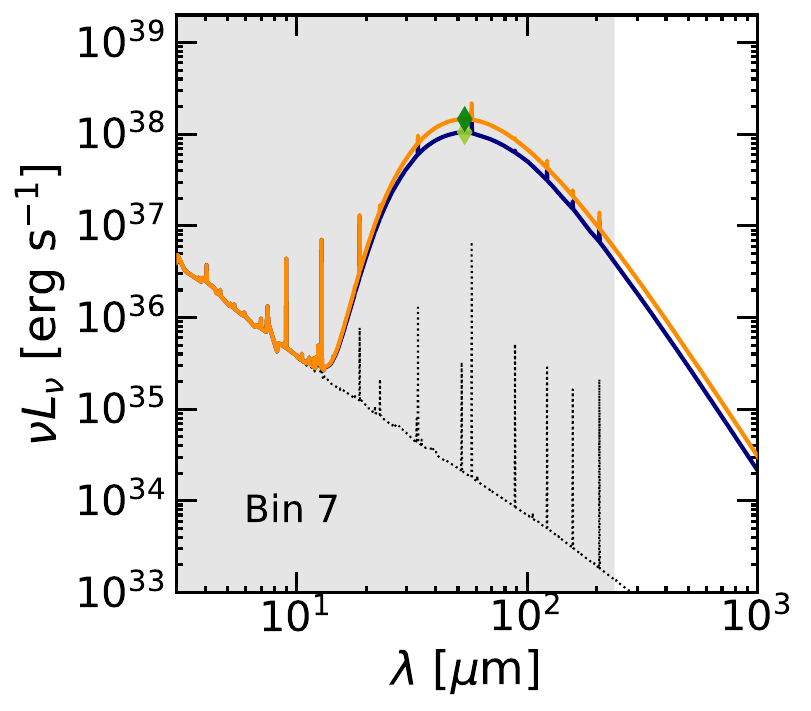}
\includegraphics[width=0.24\textwidth]{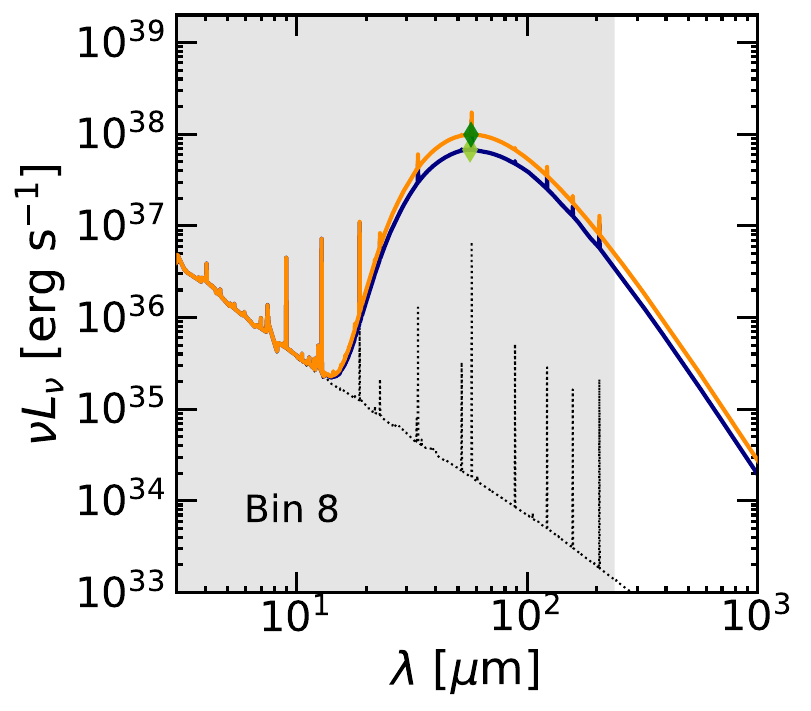}
\includegraphics[width=0.24\textwidth]{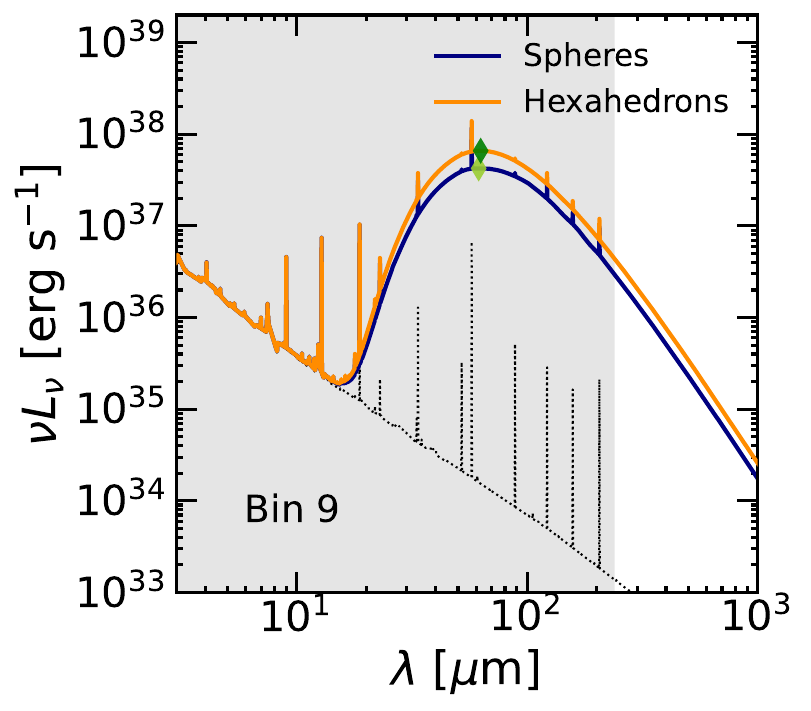}
\caption{Spectra from \textsc{cloudy} photoionization models with AC amorphous carbon grains, comparing spherical (blue line) and hexahedral-shaped (orange line) grains. Each panel shows models using only the grain sizes corresponding to bins 2 through 9 from Table~\ref{tab:bines}. The contribution of the stellar spectrum is shown in black, and green diamonds mark the IR continuum peak of each model. The shaded grey region indicates wavelengths where opacities were derived from TAMUdust2020 output data.}
\label{fig:Modelac1app}
\end{figure*}

\begin{figure*}
\centering
\includegraphics[width=0.24\textwidth]{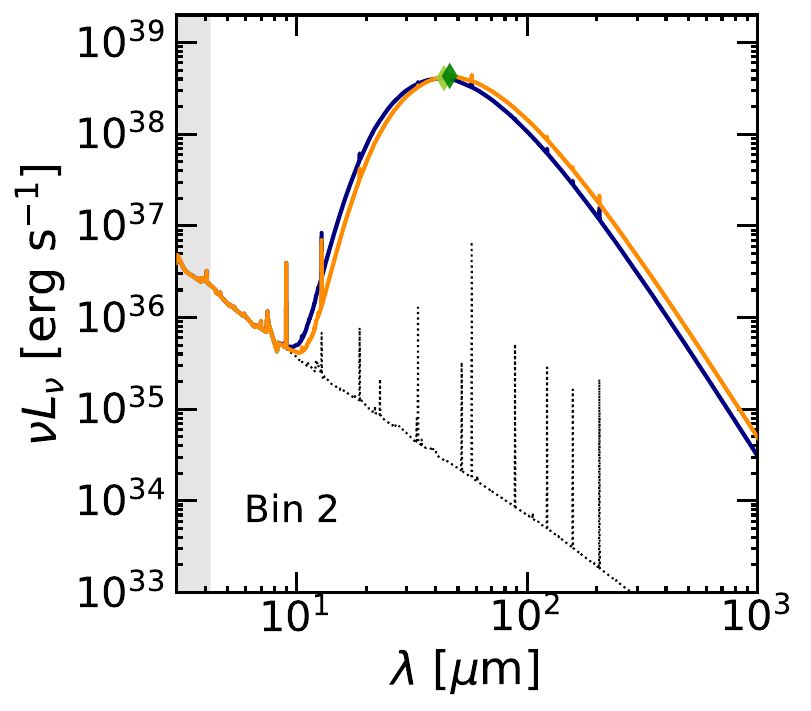}
\includegraphics[width=0.24\textwidth]{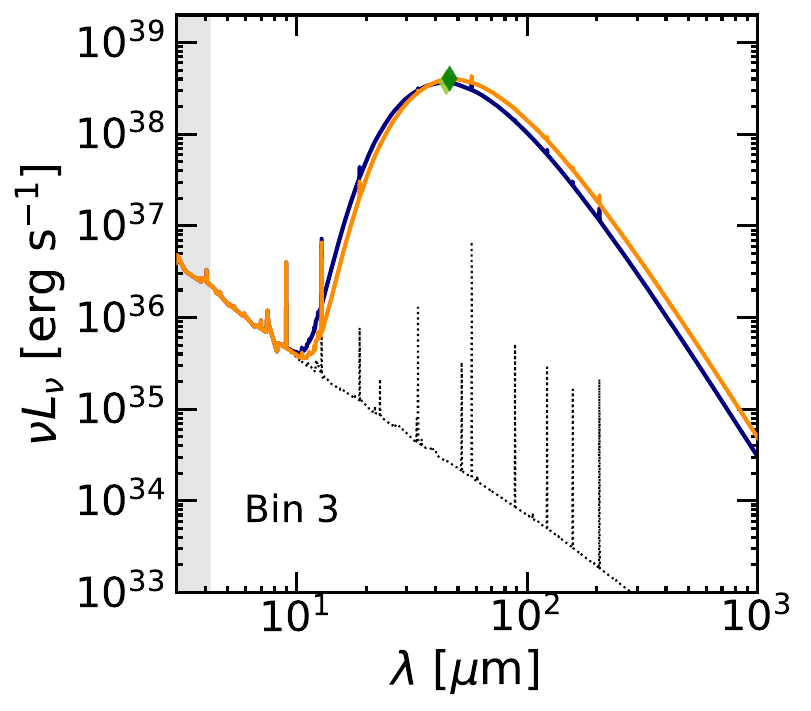}
\includegraphics[width=0.24\textwidth]{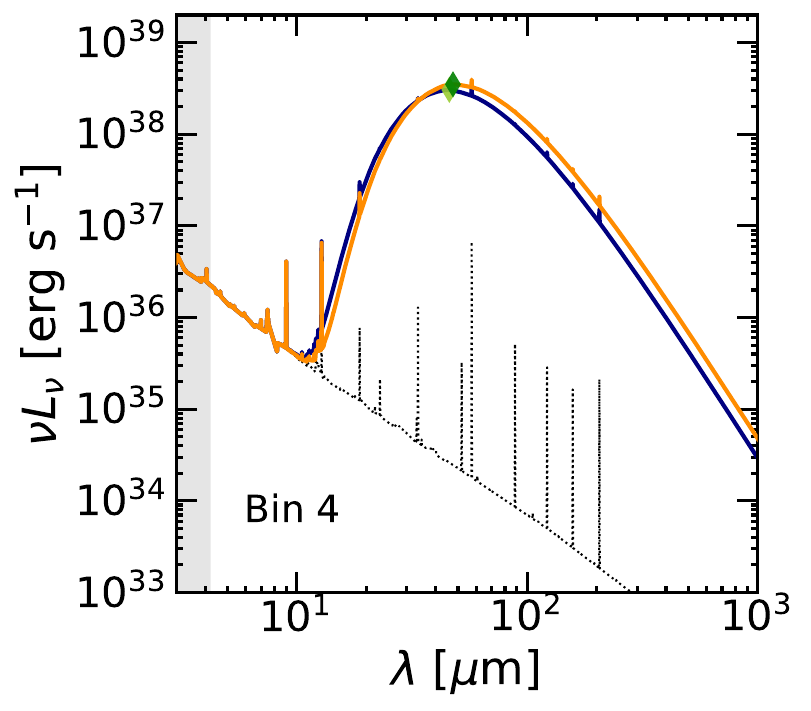}
\includegraphics[width=0.24\textwidth]{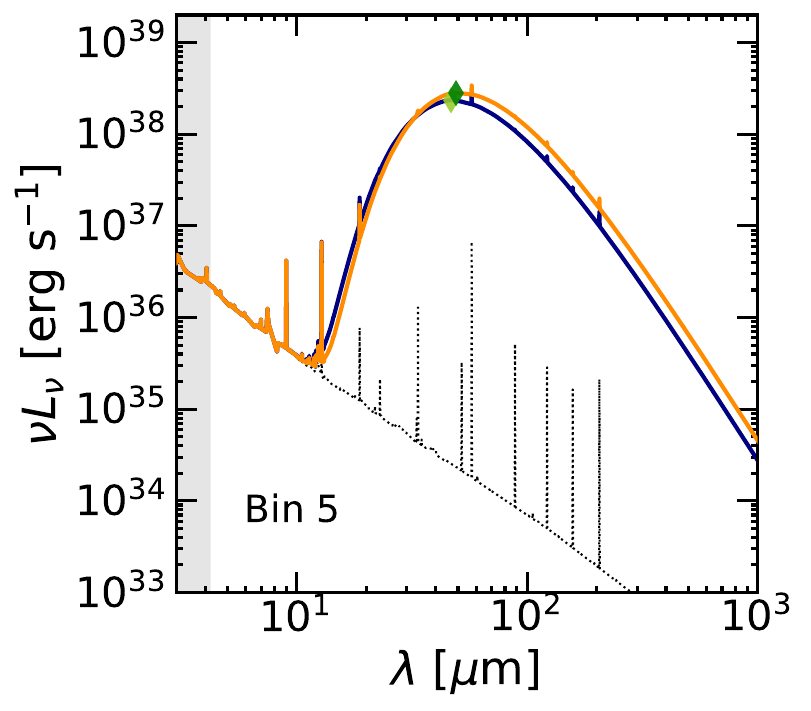}\\
\includegraphics[width=0.24\textwidth]{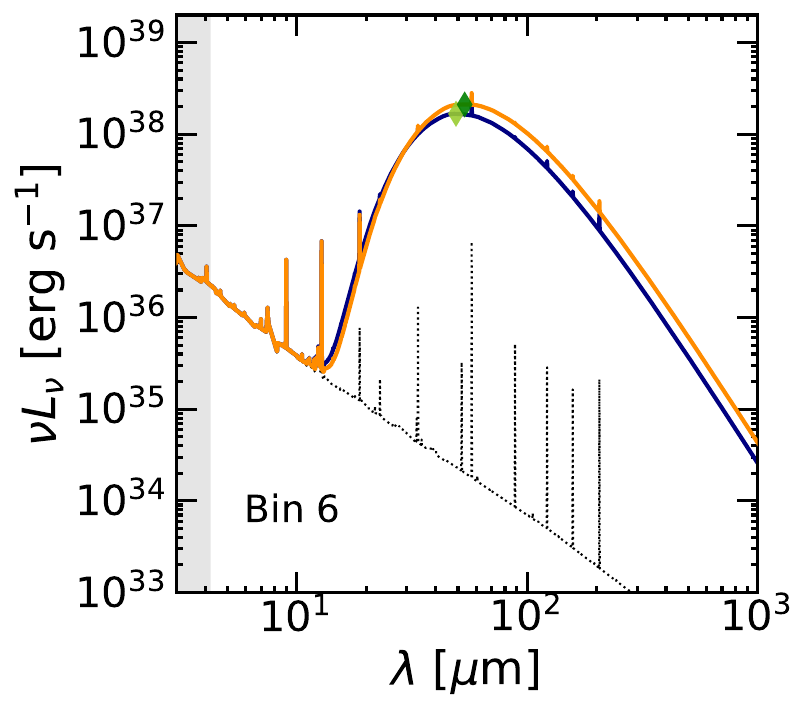}
\includegraphics[width=0.24\textwidth]{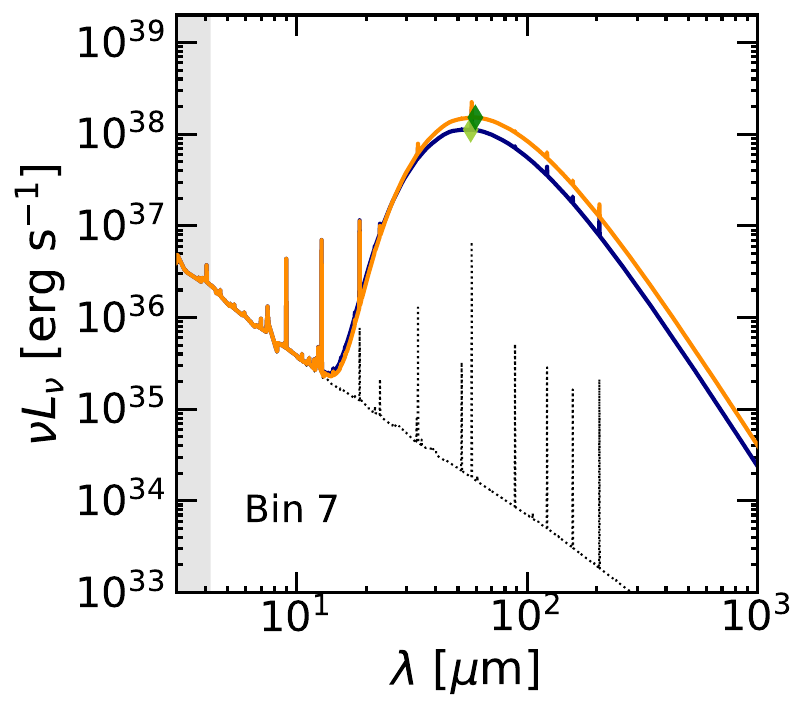}
\includegraphics[width=0.24\textwidth]{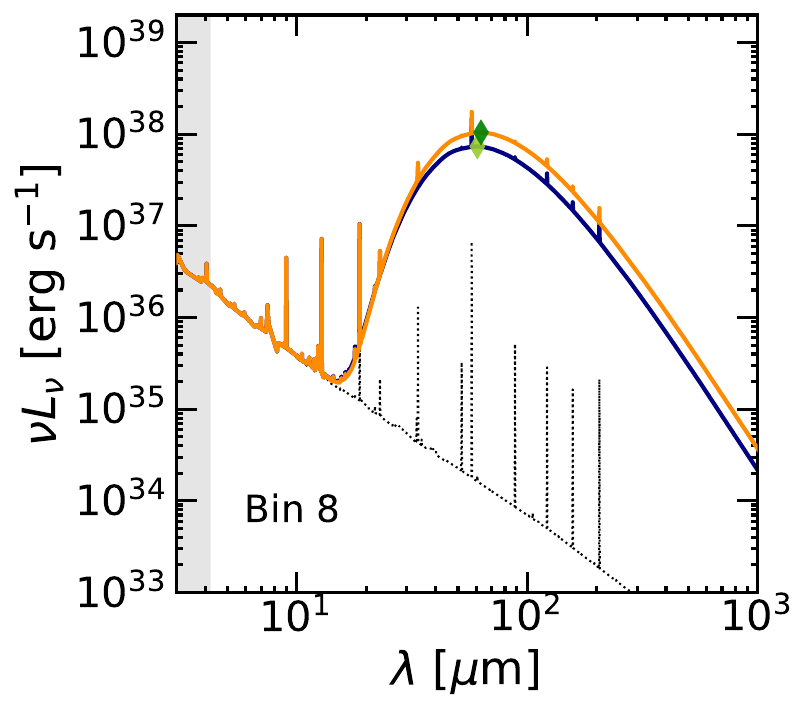}
\includegraphics[width=0.24\textwidth]{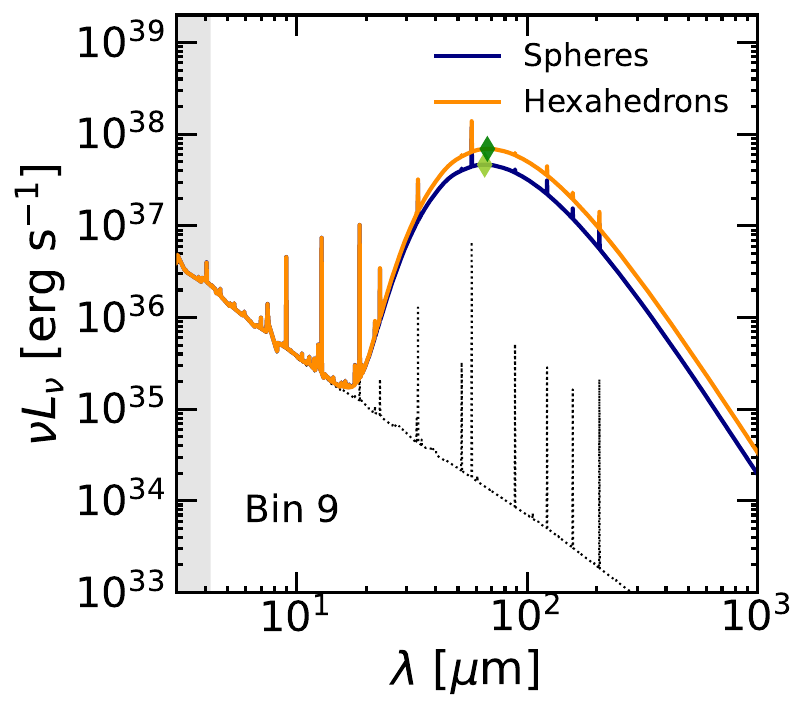}
\caption{Spectra from \textsc{cloudy} photoionization models with BE amorphous carbon grains, comparing spherical (blue line) and hexahedral-shaped (orange line) grains. Each panel shows models using only the grain sizes corresponding to bins 2 through 9 from Table~\ref{tab:bines}. The contribution of the stellar spectrum is shown in black, and green diamonds mark the IR continuum peak of each model. The shaded grey region indicates wavelengths where opacities were derived from TAMUdust2020 output data.}
\label{fig:Modelbe1app}
\end{figure*}


\bsp	
\label{lastpage}
\end{document}